\documentclass[10pt,journal,cspaper,compsoc]{IEEEtran}
\usepackage{amsmath}
\usepackage{txfonts}
\usepackage{amssymb}
\usepackage{graphicx}
\usepackage{booktabs}
\usepackage{comment}
\usepackage{subfigure}
\usepackage{extarrows}
\usepackage{multirow}
\usepackage{multicol}
\newtheorem{definition}{Definition}
\newtheorem{theorem}{Theorem}
\newtheorem{lemma}{Lemma}

\newtheorem{corollary}{Corollary}

\newtheorem{assumption}{Assumption}
\newtheorem{axiom}{Axiom}
\newtheorem{assertion}{Assertion}

\def\newblock{\hskip .11em plus .33em minus .07em}
\DeclareGraphicsExtensions{.pdf,.png,.jpg,.eps}
\usepackage[lined,ruled,vlined]{algorithm2e}

\begin{document}

  \title{A Linear Circuit Model For Social Influence}   
\author{Biao Xiang$^1$, Enhong Chen$^1$~\IEEEmembership{Senior Member,~IEEE}, Qi Liu$^1$, Hui Xiong$^2$,~\IEEEmembership{Senior Member,~IEEE}
\IEEEcompsocitemizethanks{\IEEEcompsocthanksitem Enhong Chen is with the School
of Computer Science and Technology, University of Science and Technology of China.\protect\\
E-mail: cheneh@ustc.edu.cn
}
\thanks{}}

\IEEEcompsoctitleabstractindextext{%
\begin{abstract}

Understanding the behaviors of information propagation is essential
for the effective exploitation of social influence in social
networks. However, few existing influence models are both tractable
and efficient for describing the information propagation process and
quantitatively measuring social influence. To this end, in this
paper, we develop a linear social influence model, named Circuit due
to its close relation to the circuit network. Based on the
predefined four axioms of social influence, we first demonstrate
that our model can efficiently measure the influence strength
between any pair of nodes. Along this line, an upper bound of the
node(s)' influence is identified for potential use, e.g., reducing
the search space. Furthermore, we provide the physical implication
of the Circuit model and also a deep analysis of its relationships
with the existing methods, such as PageRank. Then, we propose that
the Circuit model provides a natural solution to the problems of
computing each single node's authority and finding a set of nodes
for social influence maximization. At last, the effectiveness of the
proposed model is evaluated on the real-world data. The extensive
experimental results demonstrate that Circuit model consistently
outperforms the state-of-the-art methods and can greatly alleviate
the computation burden of the influence maximization problem.

%

\end{abstract}

}

\maketitle

\IEEEdisplaynotcompsoctitleabstractindextext
\IEEEpeerreviewmaketitle

\section{Introduction}\label{sec:introduction}
Social networks make connections among individuals. Usually, in this social paradigm, people tend to connect with their friends, family members or colleagues, which makes the connections in social networks are a kind of trust relationship. Under this relationship, if somebody do something, her friends tend to believe something is good or trustable. For example, suppose a man bought a new product and shared his pleasant experience about it on social network site, then his social friends would be likely to be influenced by his experience and may take it as an advice when they want to buy a similar product. This is a perfect effect on product marketing and information propagating. There are two obvious reasons for this. Firstly, the recommendation from one's friends is more likely to be accepted. Secondly, this effect could trigger a domino effect, e.g., if a product is adopted and shared by someone, then her friends may take it as an advice to adopt it also, then her friends' friends and so forth. This effect is so-called ``word-of-mouth'' or ``viral marketing'' effect and has been investigated for a long time~\cite{bass2004new,brown1987social,domingos2001mining,goldenberg2001talk, goldenberg2001using,mahajan1990new,richardson2002mining}. Marketing persons, news communicators are both wondering how to take advantage of this effect to improve their work on social network platform.

To this end, it is preliminary to model the influence between individuals. Influence is the effect that an individual has on the other ones when they are making decisions or behaving, the amount of which could be viewed as a probability --- roughly speaking, suppose individual A has tried $M$ things and individual $B$ tried $N$ of those things following A, then the amount of influence from individual A to B is $N/M$, which, ranged between 0 and 1, could be viewed as a probability. If we could model the influence between individuals and get its quantity, then we could take advantage of it to design the strategy for product marketing or information propagating. However, practically, we still need to model the influence of group of individuals. For example, suppose there are more than one person sharing their experiences of a product, you will be effected by their combination influence. This type of influence is subtle, which is different from a single influence or the sum of those single influences. The modeling of combination influence is very useful. When the marketing person design a viral marketing campaign, they always select more than one individuals to endorse their product, then, the influence a person received from a viral marketing campaign are usually combination influence from multiple persons.

In recent years, there has been many theoretical and empirical studies on social influence modeling. Anagnostopoulos et al.~\cite{anagnostopoulos2008influence} proposed two statistical tests to distinguish social influence from the multi-sources of correlation(i.e. homophyly, confounding and influence respectively) between the actions of friends in a social network. Goyal et al.~\cite{goyal2010learning} studied how to learn the amount of social influence between adjacent individuals. Granovetter et al.~\cite{granovetter1978threshold} proposed a model, called as Linear Threshold(LT) model to simulate the information propagation process and give the amount of social influence between any pairs, while Goldenberg et al.~\cite{goldenberg2001talk} proposed another model, called as Independent Cascade(IC) model. However, both of them are operational models and are untractable and inefficient. Under these models, you couldn't find a closed-form solution for social influence; and if you want to get the influence of an individual on the others, you have to run Monte-Carlo simulations for a sufficiently many times (e.g. 20000 times) to obtain an approximate estimate~\cite{chen2009efficient}, which is very time-consuming.

To alleviate these obstacles, in our preliminary
work~\cite{biao2012linear}, we proposed a circuit inspired linear model to describe the influence between individuals. Specifically, we adopt a two-stage strategy to achieve this goal. In the first stage, we propose a rule-based definition to model the influence between pair of single individual and obtain a closed-form solution which a probabilistic influence matrix in which any $(i,j)$th-entry is the strength of influence from node $i$ to node $j$. In the second stage, we propose a concept ``independent influence'' with which we form the formula for the influence of group of individuals. Under this model, social influence is tractable and could be computed efficiently by a fast Gauss-Seidel iteration method. In addition, we propose a upper bound to estimate the total influence of a node in the network, which could help us to identify those actually influential nodes. Finally, by exploiting the influence model and using the upper bound to select seeds, we propose a novel method to solve the well-known viral marketing problem. The experimental results demonstrate that this method outperforms the state-of-the-art algorithms both in efficiency and effectiveness.

In this paper, we further study the linear model for influence. Unlike the two-stage strategy adopted in the preliminary study, we uniformly model the influence of individual and group of individuals by an axiomatic definition, and then find its closed-form expression. Moreover, this expression could be solved by a fast Gauss-Seidel iteration method. Along this line, we further find a compact upper bound to estimate the total influence of individual or group of individuals, which is helpful to evaluate whether or not an individual or a group of individuals is influential enough. On the power of this upper bound, we solve the viral marketing problem effectively. What's more, we find that when we use the upper bound as the approximation of its real number, the problem could be solved in nearly linear time and the experimental results on variety of networks demonstrate that this method could produce a performance better than the state-of-the-art algorithms both in efficiency and effectiveness. In addition, we also seek the relationship between our model and other traditional models~(e.g. independent cascade model~\cite{goldenberg2001talk} and stochastic model~\cite{aggarwalflow}) by theoretical and empirical ways.

The rest of the paper is organized as follows. Section~\ref{sec:relatedWork} presents the latest related work about influence model and social influence maximization. In Section~\ref{sec:theory}, we propose an axiomatic definition to model the influence of a set and then deduces its closed-form expression which could be solved by a fast Gauss-Seidel iteration method. In this section, we also propose an upper bound to estimate the total influence of a set on the network. In Section~\ref{sec:deepunderstanding}, we propose the circuit simulation of the model and seek the relationship between our model and other traditional models. In Section~\ref{sec:application}, we adopt the influence model to solve the well-known social influence maximization problem and propose two novel method, i.e. \textbf{Circuit-Complete} and \textbf{Circuit-Fast}. In Section~\ref{sec:exp}, we demonstrate three claims of this paper through experiments: linear model is close related to the traditional models; upper bound is consistently close to the real total influence; and Circuit-Complete and Circuit-Fast outperform the state-of-the-art algorithms. In Section~\ref{sec:conclusion} we conclude this paper and propose several problems to be solved in the future.

\section{Related Work}\label{sec:relatedWork}

Related work can be grouped into two categories. In the first
category, we describe some  existing social influence models.
The second category includes the existing works for the
social influence maximization problem.

\textbf{Social Influence Models.}
In the literature, many studies about social influence have been published. For instance, Anagnostopoulos et
al.~\cite{anagnostopoulos2008influence} proved the existence of
social influence by statistical tests. Also, Goyal et al.~\cite{goyal2010learning}
studied how to learn the true probabilities of social influence
between individuals. In addition, there are several models to infer how the influence propagates through the network. For example, Granovetter et al.~\cite{granovetter1978threshold} proposed the Linear Threshold(LT) model to describe it, while Goldenberg et al.~\cite{goldenberg2001talk} proposed the Independent Cascade(IC) model. Since these two models are not tractable, Kimura et al.~\cite{kimura2006tractable} proposed a comparably tractable model SPM  and Aggarwal et al.~\cite{aggarwalflow} proposed a stochastic model to address this issue. Recently, Easley
et al.~\cite{easley2010networks} and Aggarwal et
al.~\cite{aggarwal2011social} summarized  and generalized many
existing studies on social influence and some other research aspects
of social networks. More importantly, they demonstrate that by
carefully study, the information exploited from social influence can
be leveraged for dealing with the real-world problems~(e.g., the
problems from markets or social security) effectively and
efficiently.

\textbf{Social Influence Maximization.}  As an application, there is an important research branch to exploit social influence for marketing, which is called as viral marketing and target at finding a small set of ``influential'' individuals~(those individuals is called as ``seed'') of the network--- giving them free samples of a product --- for triggering a cascade of influence by which friends will recommend the product to other friends, hoping the product will be adopted by a large fraction of the network.

There are many literatures which aimed at solving this problem, here we list part of them as representation. At the beginning, Domingoes and Richardson proposed this problem firstly~\cite{domingos2001mining, richardson2002mining}. Kempe et al. formulated this problem as a discrete optimization problem and they proved that the optimization problem is NP-hard, and presented a greedy approximation~(GA) algorithm which guarantees that the influence spread result is within $(1-1/e)$ of the optimal result. To address the
efficiency issue, Leskovec et al.~\cite{leskovec2007cost} presented
a "Lazy Forward" scheme~(called as \textbf{CELF optimization}) which
take advantage of the submodularity property of the influence maximization
objective to reduce the number of evaluations on the influence
spread of individuals. To address the scalability issue, Chen et al. proposed several heuristic methods includes \textbf{DegreeDiscountIC}~\cite{chen2009efficient} and \textbf{PMIA}~\cite{chen2010scalable} which uses local
arborescence structures of each individual to approximate the social
influence propagation. Wang et al.~\cite{wang2010community} presented a community-based greedy algorithm to find the Top-$K$ influential nodes. They first detect
the communities in social network and then find influential nodes
from the selected potential communities.

\section{Social Influence Modeling}\label{sec:theory}
Social influence refers to the behavioral change of individuals affected by
others in a network. Social influence is an intuitive and well-accepted phenomenon in social networks~\cite{easley2010networks}. Here, we will provide a quantitative way to measure the social influence. To facilitate the following discussion, we list the important math notations used in this paper in Table~\ref{tab:notations}.
\begin{table*}
\centering \caption{Math Notations.}\label{tab:notations}
\begin{tabular}{|l|l|}
  \hline
  Notations        & DESCRIPTION \\ \hline
  $f_{\mathcal{S}\rightarrow \mathcal{T}}$        &  the influence from a group of individuals $\mathcal{S}$ to another group of individuals $\mathcal{T}$,\\ & where $\mathcal{S}$ and $\mathcal{T}$ could be single individual.        \\ \hline 
  $\textbf{f}_\mathcal{S}=[f_{\mathcal{S}\rightarrow 1}, f_{\mathcal{S}\rightarrow 2}, ...f_{\mathcal{S}\rightarrow n}]'$        &  the influence vector from $\mathcal{S}$ to each member of $\mathcal{V}$, where $\mathcal{S}$ could be single individual.         \\ \hline    
  $\textbf{F}=[f_{ij}]_{n*n}$      & Influence matrix where $f_{ij}$ is the influence from $i$ to $j$ and equal to $f_{i\rightarrow j}$.   \\ \hline    
  $\textbf{T}=[t_{ij}]_{n*n}$    & transmission matrix where $t_{ij}$ is the transmission probability from $i$ to $j$.      \\ \hline 
  $\Lambda=diag(\lambda_1, \lambda_2,...\lambda_n)$           &Damping coefficient matrix where $\lambda_i$ is the damping coefficient of $i$.         \\ \hline           
  $\textbf{P}=[p_{ij}]_{n*n}=(\textbf{I}+\Lambda-\textbf{T}')^{-1}$ &  Basis matrix where $p_{ij}$ is the basis element to compound influence.             \\ \hline
$\textbf{p}_{\mathcal{T}}=[p_{1\rightarrow{\mathcal{T}}},p_{2\rightarrow{\mathcal{T}}},...p_{n\rightarrow\mathcal{T}}]'$ &Potential vector where $p_{j\rightarrow\mathcal{T}}=\sum_{i\in \mathcal{T}}p_{ij}$\\ \hline
$b_{\mathcal{S}\rightarrow \mathcal{T}}$ & the upper bound of $f_{\mathcal{S}\rightarrow \mathcal{T}}$, equals to $\sum_{j\in \mathcal{S}}((1+\lambda_j)-\sum_{k\in \mathcal{S}}t_{kj})p_{j\rightarrow\mathcal{T}}$\\ \hline
  $\Theta=diag(\theta_1,\theta_2,...\theta_n)$   & $\theta_i = \sum_{j=1}^{n}t_{ji}$ is the total transmission probabilities flowing into individual $i$.        \\ \hline
\end{tabular}
\end{table*}

Let $\mathcal{G}=(\mathcal{\mathcal{V}},\mathcal{E})$ be a social network, where the node set $\mathcal{V}=\{1,2,...,n\}$ includes all of individuals, the edge set $\mathcal{E}$ represents all the social connections which could be viewed as trust relationships. We denote the influence from a group of individuals $\mathcal{\mathcal{S}}$ to another group of individuals $\mathcal{T}$ as $f_{\mathcal{\mathcal{S}}\rightarrow \mathcal{T}}$, where $\mathcal{S}$ and $\mathcal{T}$ are subsets of $\mathcal{V}$. When $\mathcal{S}$ and $\mathcal{T}$ are sets with only one element, $f_{\mathcal{S}\rightarrow \mathcal{T}}$ is the influence between pair of single individuals. Under this notation, the viral marketing campaign design could be formulated as the following optimization problem
\begin{equation}
  \mathcal{S}={\arg\max}_{\mathcal{S}\subset \mathcal{V}}f_{\mathcal{S}\rightarrow \mathcal{V}}\ \ \ \ subject\ to\ \ |\mathcal{S}|=K
\end{equation}

In this paper, we propose four axioms to model the general influence $f_{\mathcal{S}\rightarrow \mathcal{T}}$ as follows.

\begin{axiom}\label{ax:1}
      The influence from $\mathcal{S}$ to $\mathcal{T}$ is equal to the sum of influence from $\mathcal{S}$ to each member of $\mathcal{T}$, that is
  \begin{equation}\label{eq:rule1}
    f_{\mathcal{S}\rightarrow \mathcal{T}}=\sum_{i\in \mathcal{T}}f_{\mathcal{S}\rightarrow i}.
  \end{equation}
\end{axiom}

\begin{axiom}\label{ax:2}
  If $i$ is a member of $\mathcal{S}$, the influence from $\mathcal{S}$ to $i$ should be always equal to 1, that is
  \begin{equation}
    \label{eq:rule2}
    f_{\mathcal{S}\rightarrow i}=1\ \ \ \ for\ i\in \mathcal{S}.
  \end{equation}
\end{axiom}

\begin{axiom}\label{ax:3}
  Influence could transmit through the trust connection in network with a certain transmission probability on it.
\end{axiom}

\begin{axiom}\label{ax:4}
  The influence to an arbitrary individual is determined by the influences to her trust-friends. Suppose $j$'s trust-friends set is $N_j=\{j_1,j_2,...j_m\}$~(i.e. $\forall k\in N_j$, there is a trust connection $(j,k)\in E$) and the influence to $k\in N_j$ is $f_{\mathcal{S}\rightarrow k}$, then
\begin{equation}\label{eq:mix}
  f_{\mathcal{S}\rightarrow j} = f_j(t_{j_1j}f_{\mathcal{S}\rightarrow j_1},\  t_{j_2j}f_{\mathcal{S}\rightarrow j_2},...\ t_{j_mj}f_{\mathcal{S}\rightarrow j_m})
  \ \ for\ j\not{\in}\mathcal{S}
\end{equation}
where $f_j(*)$ is a combination function for $j$ and $t_{kj}$ is the transmission probability on trust connection $(j,k)$~\footnote{Notably, in social networks, the direction of trust connection is inverse to the direction of influencing, which means that if $i$ trusts $j$ then $j$ will influence $i$. }.
\end{axiom}

Based on the above four axioms, there are two factors which will determine the shape of the social influence model.

The first factor is the transmission probabilities on each trust connection. In this paper, we use an assumption to confine the probability, that is
\begin{assumption}\label{as:theta}
  The sum of transmission probabilities flowing into one node should be less than or equal to 1. That is,
$$
  \theta_i = \sum_{j=1}^{n}t_{ji} \leq 1\ \ \ \ for\ \ i=1,2...n
$$
where $t_{ji}$ is the transmission probability from node $j$ to node $i$. If $(i,j)\not{\in}\mathcal{E}$, then $t_{ji}=0$.
\end{assumption}

Actually, this assumption is used for measuring the amount
of information~(e.g., with regard to an event or message) that will
be accepted by each node. The corresponding value varies in the
range of [0,1], where $0$ stands for the ignorance of the
information and $1$ means this node totally believes in it.

The second factor is the way how an individual combine the influences receiving from her trust-friends. For instance, Aggarawal et al~\cite{aggarwal2011social} proposed a way to describe this function, that is
\begin{equation}\label{eq:aggarwal}
  f_{\mathcal{S}\rightarrow j} = 1-\Pi_{k\in N_j}(1-t_{kj}f_{\mathcal{S}\rightarrow k})
\end{equation}
which claims that the transmitted influences from different friends should be independent to each other. This is a theoretically reasonable way, however it is too complex to get its closed-form solution. Thus, in this paper, we propose a linear way. That is,
\begin{equation}\label{eq:linear}
  f_{\mathcal{S}\rightarrow j} = \frac{1}{1+\lambda_j}\sum_{k\in N_j}{t_{kj}f_{\mathcal{S}\rightarrow k}} \ \ \ \ for\ j\not{\in}\mathcal{S}
\end{equation}
where $\lambda_j$ is the \textbf{damping coefficient} of $j$ for the
influence propagating. It locates in range $(0,+\infty)$. The smaller $\lambda_j$ is (i.e., approaching to 0), the less
the information will be blocked by node $j$. In real applications,
this number may also varies from the topics of the propagating
information. For instance, if node $j$ favors the topic of the
propagating information, $\lambda_j$ will approach to 0,
otherwise, it will approach to a big positive number, even $+\infty$.

\subsection{The Deduction of Influence}\label{subsec:deduction}
For Equation~\ref{eq:linear} only describes for the ones not in $\mathcal{S}$, we first reform it to describe all individuals, including the member of $\mathcal{S}$, as follows.
\begin{equation}\label{eq:linearCor}
  f_{\mathcal{S}\rightarrow i} = \frac{1}{1+\lambda_i}\sum_{j\in N_i}{(t_{ji}f_{\mathcal{S}\rightarrow j}+\nu_{\mathcal{S},i})}\ \ \ for\ i=1,2,...n
\end{equation}
where $\nu_{\mathcal{S},i}$ is a correction to guarantee that, $f_{\mathcal{S}\rightarrow i}$ is equal to 1 if $i$ is a member of $\mathcal{S}$ and $\frac{1}{1+\lambda_j}\sum_{k\in N_j}{t_{kj}f_{\mathcal{S}\rightarrow k}}$ otherwise~(Axiom~\ref{ax:2} and Axiom~\ref{ax:4}). Thus, the value of $\nu_{\mathcal{S},i}$ could be determined as
\begin{equation}\label{eq:nu}
\nu_{\mathcal{S},i}=\left\{\begin{aligned}
  &a\ number\ to\ guarantee\ f_{\mathcal{S}\rightarrow i}=1 & \ \  &i\in \mathcal{S}& \\
  &0&\ \ \ &i\not\in \mathcal{S}&
\end{aligned}\right.
\end{equation}

Equation~\ref{eq:linearCor} for $i=1,2,...n$ could be rewritten as
\begin{equation}
  \textbf{f}_\mathcal{S}=(\textbf{I}+\Lambda)^{-1}(\textbf{T}'\textbf{f}_\mathcal{S}+\nuup_\mathcal{S})\nonumber
\end{equation}
where
\begin{eqnarray}
  \textbf{f}_\mathcal{S}&=&[f_{\mathcal{S}\rightarrow 1}, f_{\mathcal{S}\rightarrow 2}, ...f_{\mathcal{S}\rightarrow n}]'\nonumber\\
  \textbf{T}&=&[t_{ij}]_{n*n}\nonumber\\
  \nuup_\mathcal{S}&=&[\nu_{\mathcal{S}, 1}, \nu_{\mathcal{S},2}, ...\nu_{\mathcal{S},n}]'\nonumber\\
  \Lambda&=&diag(\lambda_1, \lambda_2, ...\lambda_n)\nonumber
\end{eqnarray}
which could be solved as
\begin{eqnarray}
  \label{eq:closedsolution}
  \textbf{f}_\mathcal{S}&=&(\textbf{I}+\Lambda-\textbf{T}')^{-1}\nuup_\mathcal{S}\\
     &=&\textbf{P}\cdot\nuup_\mathcal{S}\label{eq:inf}
\end{eqnarray}
where the transpose of $(\textbf{I}+\Lambda-\textbf{T}')$ is strictly diagonally dominant, thus it is invertible. In this paper, we denote the inverse of $(\textbf{I}+\Lambda-\textbf{T}')$ as $\textbf{P}=[p_{ij}]_{n*n}$ and call it as \textbf{basis matrix}.

Based on Axiom 2~(the influence from $\mathcal{S}$ to the member of $\mathcal{S}$ should be 1) and Equation~\ref{eq:nu}, from Equation~\ref{eq:inf} we can get
\begin{equation}\label{eq:nuup}
f_{\mathcal{S}\rightarrow i}=\sum_{j \in \mathcal{S}}p_{ij}\nu_{\mathcal{S},j} = 1\ \ \ for\ i\in \mathcal{S}
\end{equation}
Suppose $\mathcal{S}=\{s_1,s_2,...s_K\}$ where $K$ is the cardinality of $S$, and without loss of generality we assume $s_1<s_2<...<s_K$. After denoting $\nuup_{\mathcal{S}\mathcal{S}}=[\nu_{\mathcal{S},s_1}, \nu_{\mathcal{S},s_2}, ...,\nu_{\mathcal{S},s_{K}}]'$, and denoting $\textbf{P}_{\mathcal{S}\mathcal{S}}$ as the matrix which is cut down from $\textbf{P}$ by removing the columns and rows not corresponding to members of $\mathcal{S}$. Equation~\ref{eq:nuup} could be rewritten as
$$
\textbf{P}_{\mathcal{S}\mathcal{S}}\nuup_{\mathcal{S}\mathcal{S}}=\mathbf{e}
$$
where $\mathbf{e}$ is a $|\mathcal{S}|$-dimensions vector with all 1s.  Thus,
\begin{equation}\label{eq:nuS}
  \nuup_{\mathcal{S}\mathcal{S}} = \textbf{P}_{\mathcal{S}\mathcal{S}}^{-1}\mathbf{e}
\end{equation}
and
$$
\nu_{\mathcal{S},{s_i}}=[\textbf{P}_{\mathcal{S}\mathcal{S}}^{-1}\mathbf{e}]_i
$$
Conclusively, we could form the closed-form solution of $\textbf{f}_\mathcal{S}$ as follows.
\begin{theorem}\label{th:solution}
  In a network $\mathcal{G}(\mathcal{V},\mathcal{E})$, given the transmission matrix $\textbf{T}$ and information's damping coefficient matrix $\Lambda$, the influence vector from a set $\mathcal{S}=\{s_1,s_2,...s_K\}\in \mathcal{V}$~(assuming $s_1<s_2<...<s_K$) to members of the network will be
\begin{equation}\label{eq:solution}
  \textbf{f}_\mathcal{S} = (\textbf{I}+\Lambda-\textbf{T}')^{-1}\nuup_\mathcal{S}=\textbf{P}\nuup_\mathcal{S}
\end{equation}
where $\nuup_\mathcal{S}=[\nu_{\mathcal{S},1},\nu_{\mathcal{S},2},...\nu_{\mathcal{S},n}]'$ and
\begin{equation}\label{eq:nusolution}
\nu_{\mathcal{S},i}=\left\{\begin{aligned}
  &[\textbf{P}_{\mathcal{S}\mathcal{S}}^{-1}\textbf{e}]_k& \ \ \  &i=s_k\in \mathcal{S}& \\
  &0&\ \ \ &i\not\in \mathcal{S}&
\end{aligned}\right.
\end{equation}
where $\textbf{P}_{\mathcal{S}\mathcal{S}}$ is the matrix which is cut down from $\textbf{P}$ by removing the columns and rows not corresponding to members of $\mathcal{S}$.
\end{theorem}

In other forms,
\begin{equation}\label{eq:FSlinear}
  \textbf{f}_\mathcal{S} = \sum_{i\in \mathcal{S}}\nu_{\mathcal{S},i} \textbf{P}_{\cdot i}
\end{equation}
since the $\nu_{\mathcal{S},i}$ is equal to 0 if $i\not{in}\mathcal{S}$. From this equation, we could observe that $\textbf{f}_\mathcal{S}$ is actually a linear combination of the columns of $\textbf{P}$, that is the reason why we call $\textbf{P}$ as basis matrix.

Specifically, when $\mathcal{S}$ contains only one element, let it is $i$,
\begin{equation}\label{eq:infFi}
\textbf{f}_\mathcal{S} = \nu_{\{i\},i} \textbf{P}_{\cdot i}\triangleq \textbf{f}_i
\end{equation}
and thus
\begin{equation}\label{eq:fij}
  f_{i\rightarrow j}=\nu_{\{i\},i} p_{ji}
\end{equation}
For $|\mathcal{S}|=1$, $\textbf{P}_{\mathcal{S}\mathcal{S}}$ is a $1\times 1$ matrix and equals to $[p_{ii}]$. Easily, based on Equation~\ref{eq:nusolution}, we could get
\begin{equation}\label{eq:p_ii}
\nu_{\{i\},i} =[\textbf{P}_{\mathcal{S}\mathcal{S}}^{-1}\mathbf{e}]_1=\frac{1}{p_{ii}}
\end{equation}
Equation~\ref{eq:infFi} for $i=1,2,...n$ could be rewritten as
\begin{eqnarray}\label{eq:influenceMatrix}
  \textbf{F}&\triangleq&[\textbf{f}_1, \textbf{f}_2, ...\textbf{f}_n]'\nonumber\\
   &=&[\frac{1}{p_{11}}\textbf{P}_{\cdot 1},\frac{1}{p_{22}}\textbf{P}_{\cdot 2},...\frac{1}{p_{nn}}\textbf{P}_{\cdot n}]'\nonumber\\
   &=&diag(\textbf{P})^{-1}\textbf{P}'
\end{eqnarray}
where the $(i,j)$-entry of $\textbf{F}$ is the influence from $i$ to $j$. Thus, we call $\textbf{F}=[f_{ij}]_{n*n}=[\frac{p_{ji}}{p_{ii}}]_{n*n}$ as the \textbf{influence matrix} of $\mathcal{G}$. $\textbf{F}$ gives all the influences between any pair of individuals. Given $\textbf{F}$, if want to know the influence from $i$ to $j$, we only need to look up the value at the $(i,j)$-entry of $\textbf{F}$, that is $f_{i\rightarrow j}=f_{ij}=\frac{p_{ji}}{p_{ii}}$.

\subsection{The Computation of $\textbf{f}_\mathcal{S}$}\label{subsec:computation}
It seems that, to compute the influence vector $\textbf{f}_\mathcal{S}$, it should compute two inverse matrices, $(\textbf{I}+\Lambda-\textbf{T}')^{-1}$ and $\textbf{P}_{\mathcal{S}\mathcal{S}}^{-1}$, thus the time complexity of this computation should be $O(n^3)$. But, fortunately, based on Equation~\ref{eq:FSlinear}, we only need to compute the columns of $\textbf{P}$ corresponding to the members of $\mathcal{S}$. Moveover, because the transpose of $(\textbf{I}+\Lambda-\textbf{T}')$ is a strictly diagonally dominant matrix, it satisfies the convergence condition of Gauss-Seidel method, it's inverse could be computed in a very fast way through a Gauss-Seidel iteration process.

Because $\textbf{P}$ is the inverse of $(\textbf{I}+\Lambda-\textbf{T}')$, there is
$$
(\textbf{I}+\Lambda-\textbf{T}') \textbf{P}_{\cdot i}=\mathbf{e}_i,
$$
where $\textbf{P}_{\cdot i}$ could be viewed as the variables of this linear system of equations. For the transpose of $(\textbf{I}+\Lambda-\textbf{T}')$ is strictly diagonally dominant, $\textbf{P}_{\cdot i}$ could be solved by Gauss-Seidel method. Specifically, Gauss-Seidel method is an iterative method which is operated as the following procedures:
\begin{enumerate}
  \item[1.] Set $p_{ji}^{(0)}=0$ for $j = 1,2,...n$;
  \item[2.] $p_{ji}^{(k+1)}=\frac{1}{1+\lambda_j}(\mathbf{e}_{ij}+\sum_{l>j}t_{lj}p_{li}^{(k)}+\sum_{l<j}t_{lj}p_{li}^{(k+1)})$, for $j=1, 2, ...n$;
  \item[3.] continue Step 2 until the changes made by an iteration are below certain tolerance.
\end{enumerate}
This procedures is efficient. To get $\textbf{P}_{\cdot i}$ within a valid tolerance range, it often need only dozens of iterations. Thus, the time complexity of computing $|\mathcal{S}|$ columns of $\textbf{P}$ is $O(|\mathcal{S}||E|)$. Notably, $\mathcal{S}$ is often a set with a small amount of elements, then the computation of $\textbf{P}_{\mathcal{S}\mathcal{S}}^{-1}$ only consumes constant time. Additionally, in the following sections, we will propose a method to compute $\textbf{f}_\mathcal{S}$ in $O(|E|)$ no matter how many the carnality of $\mathcal{S}$ is.

\subsection{An Upper Bound Of $f_{\mathcal{S}\rightarrow \mathcal{T}}$}\label{subsec:bound}
Based on Axiom~\ref{ax:1}, we know that $f_{\mathcal{S}\rightarrow \mathcal{T}}$ is equal to the sum of influence from $\mathcal{S}$ to each member of $\mathcal{T}$. If we define
\begin{equation}\label{eq:potential}
  p_{i\rightarrow\mathcal{T}} = \sum_{j\in \mathcal{T}}p_{ji}
\end{equation}
, we could get
\begin{theorem}
  \label{th:S2Vbound} The amount of influence from a group of individuals $\mathcal{S}$ to another group of individuals $\mathcal{T}$ has an upper bound, that is
\begin{equation}\label{eq:S2Vboundbest}
    f_{\mathcal{S}\rightarrow \mathcal{T}}\leq\sum_{i\in \mathcal{S}}((1+\lambda_i)-\sum_{k\in \mathcal{S}}t_{ki})p_{i\rightarrow\mathcal{T}}
\end{equation}
\end{theorem}
To prove this theorem, let's first prove a lemma about the correction vector $\nuup_\mathcal{S}$.
\begin{lemma}
  \label{lm:nuBound}The correction vector $\nuup_\mathcal{S}$ satisfies
  \begin{equation}\label{neq:nubound}
  \nu_{\mathcal{S},j} \leq (1+\lambda_j)-\sum_{k\in \mathcal{S}}t_{kj}\ \ \ for\ j\in \mathcal{S}
  \end{equation}
  \begin{proof}
    First, let's denote
$$
   \Gamma=(\textbf{I}+\Lambda-\textbf{T}')=\left [\begin{aligned}
   &\Gamma_{\mathcal{S}\mathcal{S}}&\ &\Gamma_{\mathcal{S}\overline{\mathcal{S}}}&\\
   &\Gamma_{\overline{\mathcal{S}}\mathcal{S}}&\ &\Gamma_{\overline{\mathcal{S}}\overline{\mathcal{S}}}&
  \end{aligned}\right ]
$$
where we rearrange and divide $\Gamma=(\textbf{I}+\Lambda-\textbf{T})$ into four submatrices based on whether or not the row's or column's corresponding individual is a member of set $\mathcal{S}$.
    From the linear algebra theory, we have
    \begin{eqnarray}\label{eq:partMatrix}
    \textbf{P} &=& \left [\begin{aligned}
   &\textbf{P}_{\mathcal{S}\mathcal{S}}&\ &\textbf{P}_{\mathcal{S}\overline{\mathcal{S}}}&\\
   &\textbf{P}_{\overline{\mathcal{S}}\mathcal{S}}&\ &\textbf{P}_{\overline{\mathcal{S}}\overline{\mathcal{S}}}&
  \end{aligned}\right ] =\Gamma^{-1}=\left [\begin{aligned}
   &\Gamma_{\mathcal{S}\mathcal{S}}&\ &\Gamma_{\mathcal{S}\overline{\mathcal{S}}}&\\
   &\Gamma_{\overline{\mathcal{S}}\mathcal{S}}&\ &\Gamma_{\overline{\mathcal{S}}\overline{\mathcal{S}}}&
  \end{aligned}\right ]^{-1}\nonumber\\
  &=&\left [\begin{aligned}
   &\textbf{M}&\ &-\textbf{M}\Gamma_{\mathcal{S}\overline{\mathcal{S}}}{\Gamma_{\overline{\mathcal{S}}\overline{\mathcal{S}}}^{-1}}&\\
   &-{\Gamma_{\overline{\mathcal{S}}\overline{\mathcal{S}}}^{-1}}\Gamma_{\overline{\mathcal{S}}\mathcal{S}}\textbf{M}&\ &{\Gamma_{\overline{\mathcal{S}}\overline{\mathcal{S}}}^{-1}}+{\Gamma_{\overline{\mathcal{S}}\overline{\mathcal{S}}}^{-1}}\Gamma_{\overline{\mathcal{S}}\mathcal{S}}\textbf{M}\Gamma_{\mathcal{S}\overline{\mathcal{S}}}{\Gamma_{\overline{\mathcal{S}}\overline{\mathcal{S}}}^{-1}}&
  \end{aligned}\right ]\nonumber
    \end{eqnarray}
    where
$$
    \textbf{M} = (\Gamma_{\mathcal{S}\mathcal{S}}-\Gamma_{\mathcal{S}\overline{\mathcal{S}}}\Gamma_{\overline{\mathcal{S}}\overline{\mathcal{S}}}^{-1}\Gamma_{\overline{\mathcal{S}}\mathcal{S}})^{-1}.
$$
    Thus, $\textbf{P}_{\mathcal{S}\mathcal{S}}=\textbf{M}$ and there is
$$
\nuup_{\mathcal{S}\mathcal{S}}=\textbf{P}_{\mathcal{S}\mathcal{S}}^{-1}\mathbf{e}=\Gamma_{\mathcal{S}\mathcal{S}}\mathbf{e}-\Gamma_{\mathcal{S}\overline{\mathcal{S}}}\Gamma_{\overline{\mathcal{S}}\overline{\mathcal{S}}}^{-1}\Gamma_{\overline{\mathcal{S}}\mathcal{S}}\mathbf{e}.
$$
Because $\Gamma_{\mathcal{S}\overline{\mathcal{S}}}\Gamma_{\overline{\mathcal{S}}\overline{\mathcal{S}}}^{-1}\Gamma_{\overline{\mathcal{S}}\mathcal{S}}$ is a nonnegative matrix~\footnote{where $\Gamma_{\overline{\mathcal{S}}\overline{\mathcal{S}}}$ is a strictly diagonal dominant matrix, thus its inverse~(denoted as $\textbf{N}=[n_{ij}]$) is a nonnegative matrix. Let's denote $\textbf{K}=\Gamma_{\mathcal{S}\overline{\mathcal{S}}}\textbf{N}\Gamma_{\overline{\mathcal{S}}\mathcal{S}}=[k_{ij}]$, there is $k_{ij}=\sum_{l\not{\in}\mathcal{S}}\sum_{m\not{\in}\mathcal{S}}(\gamma_{il}n_{lm}\gamma_{mj})$. Because $\gamma_{il}=-t_{il}\leq 0$, $\gamma_{mj}=-t_{mj}\leq 0$, and $n_{lm}\geq 0$, $k_{ij}\geq 0$. Thus, $\textbf{K}=\Gamma_{\mathcal{S}\overline{\mathcal{S}}}\Gamma_{\overline{\mathcal{S}}\overline{\mathcal{S}}}^{-1}\Gamma_{\overline{\mathcal{S}}\mathcal{S}}$ is a nonnegative matrix.}, we can get
$$
\nuup_{\mathcal{S}\mathcal{S}}\leq\Gamma_{\mathcal{S}\mathcal{S}}\mathbf{e}
$$
From this inequality, we can get, when $j\in \mathcal{S}$,
$$
  \label{neq:nu}
  \nu_{\mathcal{S},j}\leq (1+\lambda_j)-\sum_{k\in \mathcal{S}}t_{kj}
$$
  \end{proof}
\end{lemma}

Then, from Equation~\ref{eq:FSlinear}, based on Lemma~\ref{lm:nuBound}, there is
$$
\textbf{f}_\mathcal{S} = \sum_{i\in \mathcal{S}}\nu_{\mathcal{S},i} \textbf{P}_{\cdot i}\leq \sum_{i\in \mathcal{S}}((1+\lambda_i)-\sum_{k\in \mathcal{S}}t_{ki})\textbf{P}_{\cdot i}
$$
Thus, $f_{\mathcal{S}\rightarrow j}\leq \sum_{i\in \mathcal{S}}((1+\lambda_i)-\sum_{k\in \mathcal{S}}t_{ki})p_{ji}$, and then
\begin{eqnarray}
  f_{\mathcal{S}\rightarrow \mathcal{T}}&=&\sum_{j\in \mathcal{T}}f_{\mathcal{S}\rightarrow j}\nonumber\\
  &\leq&\sum_{j\in \mathcal{T}}\sum_{i\in \mathcal{S}}((1+\lambda_i)-\sum_{k\in \mathcal{S}}t_{ki})p_{ji}\nonumber\\
  &=&\sum_{i\in \mathcal{S}}((1+\lambda_i)-\sum_{k\in \mathcal{S}}t_{ki})p_{i\rightarrow\mathcal{T}}\nonumber
\end{eqnarray}
Thus, Theorem~\ref{th:S2Vbound} is proved.

\indent \textbf{Discussion.} Let's denote $\textbf{p}_{\mathcal{T}} = [p_{1\rightarrow\mathcal{T}},p_{2\rightarrow\mathcal{T}},...p_{n\rightarrow\mathcal{T}}]'$, $\textbf{p}_{\mathcal{T}}$ is a quantity that can be computed in $O(|E|)$ time. Because
$$
p_{i\rightarrow\mathcal{T}} = \sum_{j\in \mathcal{T}}p_{ji}=\textbf{P}_{\cdot i}'\mathbf{e}_{\mathcal{T}}\ \ \ for\ i=1,2,...,n
$$
where $\mathbf{e}_{\mathcal{T}}=[e_1,e_2,...e_n]'$, $e_i$ is equal to 1 if $i$ is a member of $\mathcal{T}$ and 0 otherwise, thus
$$
\textbf{p}_{\mathcal{T}} = \textbf{P}'\mathbf{e}_{\mathcal{T}}.
$$
Then,
$$
(\textbf{P}')^{-1}\textbf{p}_{\mathcal{T}}=(\textbf{I}+\Lambda-\textbf{T})\textbf{p}_{\mathcal{T}}=\mathbf{e}_{\mathcal{T}}
$$
which is a linear system of equations with variance $\textbf{p}_{\mathcal{T}}$ and could be solved by Gauss-Seidel method for $(\textbf{I}+\Lambda-\textbf{T})$ is a strictly diagonally dominant matrix. Thus, we could compute $\textbf{p}_{\mathcal{T}}$ in $O(|E|)$ time by the procedures similar to in Section~\ref{subsec:computation}.
Thus, if we spend $O(|E|)$ time to get $\textbf{p}_{\mathcal{T}}$ first, then the upper bound of influence from $\mathcal{S}$ to $\mathcal{T}$ will be a number could be got instantly. Moreover, because this upper bound proposed in Theorem~\ref{th:S2Vbound} is actually very consistently close to the real $f_{\mathcal{S}\rightarrow \mathcal{T}}$~\footnote{which will be verified in the experimental part of this paper}, in this paper we denote
\begin{equation}\label{eq:bound}
  b_{\mathcal{S}\rightarrow \mathcal{T}}=\sum_{j\in \mathcal{S}}((1+\lambda_j)-\sum_{k\in \mathcal{S}}t_{kj})p_{j\rightarrow\mathcal{T}}
\end{equation}
and often use it to substitute for the real $f_{\mathcal{S}\rightarrow \mathcal{T}}$ if necessary.

As a consequence of Theorem~\ref{th:S2Vbound}, we could get the following important corollary
\begin{corollary}\label{cor:fiVbound}
  \begin{equation}
  f_{i\rightarrow \mathcal{T}}\leq (1+\lambda_i)p_{i\rightarrow\mathcal{T}}=b_{i\rightarrow \mathcal{T}}
\end{equation}
\end{corollary}

\section{Deep Understandings}\label{sec:deepunderstanding}

\subsection{Another Deduction for Influence And A Physical Implication}\label{subsec:anotherdeduction}
In Section~\ref{subsec:deduction}, we proposed a way to rewrite the formula of influence and get its closed-form expression. In this section, we will propose another way to rewrite that formula and get another closed-form expression of it. But in essence, the two expressions is equivalent to each other.

Equation~\ref{eq:linear} could be rewritten as
$$
f_{\mathcal{S}\rightarrow j} = \frac{1}{1+\lambda_j}\sum_{k\not{\in}\mathcal{S}}{t_{kj}f _{\mathcal{S}\rightarrow k}}+\frac{1}{1+\lambda_j}\sum_{k\in \mathcal{S}}t_{kj} \ \ \ \ for\ j\not{\in}\mathcal{S}
$$
since $f_{\mathcal{S}\rightarrow k}=1$ if $k\in \mathcal{S}$~(Axiom~\ref{ax:2}). This equation is equivalent to
$$
(1+\lambda_j)f_{\mathcal{S}\rightarrow j} - \sum_{k\not{\in}\mathcal{S}}{t_{kj}f_{\mathcal{S}\rightarrow k}}=\sum_{k\in \mathcal{S}}t_{kj} \ \ \ \ for\ j\not{\in}\mathcal{S}
$$
which could be rewritten as
\begin{equation}\label{eq:anotherdeduction}
  (\textbf{I}+\Lambda_{\overline{\mathcal{S}\mathcal{S}}}-\textbf{T}_{\overline{\mathcal{S}\mathcal{S}}})\textbf{f}_{\overline{\mathcal{S}}}=\mathbf{t}_{\overline{\mathcal{S}}}
\end{equation}
where $\Lambda_{\overline{\mathcal{S}\mathcal{S}}}$ and $\textbf{T}_{\overline{\mathcal{S}\mathcal{S}}}$ are the matrices cut down from $\Lambda$ and $\textbf{T}$ by removing the columns and rows corresponding to the members of $\mathcal{S}$ respectively, $\textbf{f}_{\overline{\mathcal{S}}}$ is the vector cut down from $\textbf{f}_\mathcal{S}$ by removing the entries corresponding to the members of $\mathcal{S}$, and
$$
\mathbf{t}_{\overline{\mathcal{S}}}=[\sum_{k\in \mathcal{S}}t_{ki}]_{(n-|\mathcal{S}|)*1}\ \ \ \ for \ i\not{\in}\mathcal{S}
$$
For $(\textbf{I}+\Lambda_{\overline{\mathcal{S}\mathcal{S}}} - \textbf{T}_{\overline{\mathcal{S}\mathcal{S}}})$ is still a strictly diagonally dominant matrix, this linear equation system could be solved by Gauss-Seidel method in $O(|E|)$ time. That's means, we could get $\textbf{f}_{\overline{\mathcal{S}}}$~(and thus $\textbf{f}_\mathcal{S}$ also) in $O(|E|)$ time.

\begin{figure}[!h]
  \begin{center}\hspace{-1cm}
  \includegraphics[scale=0.33]{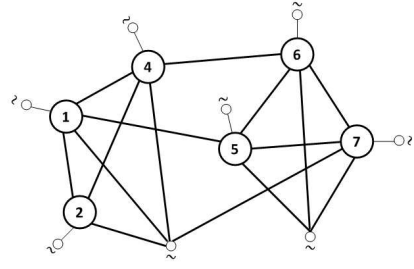} \caption{The Another Circuit Network. \label{fig:cn2}}
  \end{center}
\end{figure}

Interestingly, this deduction of influence has a circuit implication for undirected network $\mathcal{G}$. If we construct the circuit network as follows
\begin{itemize}
  \item First, construct a topologically isomorphic circuit network of $\mathcal{G}$, where the conductance between $i$ and $j$ is equal to the weight $c_{ij}$ of trust relationship $(i,j)$ (If $(i, j)$ does not exist, $c_{ij}=0$) and guarantees that $\frac{c_{ij}}{d_j}=\frac{t_{ij}}{\theta_j}$ where $d_j=\sum_{i=1}^{n}{c_{ij}}$;
  \item Second, connect $i\not{\in}\mathcal{S}$ with an external electrode $\tilde{E_i}$ through an additional electric conductor with conductance $\frac{(1+\lambda_i-\theta_i)d_i}{\theta_i}$. The electric potential value on $\tilde{E_i}$ is always $0$.
  \item Third, put a electrode pole on each $j\in \mathcal{S}$ with potential value $1$.
\end{itemize}
the potential values on the circuit network~(illustrated in Figure~\ref{fig:cn2}) will be equal to the social influence $\textbf{f}_{\mathcal{S}}$. This could be verified quite easily: for each member $i$ of $\mathcal{S}$, because there is a electrode pole with potential value $1$ on it, its potential value will be always $1$ which is equal to $f_{\mathcal{S}\rightarrow i}$; for $i\not{\in}\mathcal{S}$, based on Kirchhoff equations~\cite{kirchhoff}, there is
\begin{equation}
\sum{I}_{i}=\sum_{j=1}^n{c_{ji}(\tilde{U}_{j}-\tilde{U}_{i})}+\frac{(1+\lambda_i-\theta_i)d_i}{\theta_i}
(0-\tilde{U}_{i})=0\ \ \ for\ i\not{\in}\mathcal{S}
\end{equation}
and this equation could be reformed as
$$
\tilde{U}_i=\frac{1}{1+\lambda_i}\sum_{i=1}^n t_{ji}\tilde{U}_j\ \ for\ i\not{in}\mathcal{S}
$$
which is equivalent to the Equation~\ref{eq:linear}. Thus, the potential values on the circuit network will be equivalent to social influence vector $\textbf{f}_{\mathcal{S}}$.

\subsection{The Relationship Between Linear Model And Traditional Influence Models}\label{subsec:relation2tranditionalmodels}
In this section, we will discuss the relationship between linear model and the other models to verify the rationality of linear model.

\subsubsection{Relationship with Independent Cascade Model} Independent Cascade(IC) model is a well known and mostly-studied influence model. Under this model, if individual $i$ is activated at time $t$, then it will influence her each not-yet-activated friend at time $t+1$ (and only at time $t+1$) with a transition probability, until no new individual is activated. Although IC model has been mostly studied, its inefficiency is always a serious drawback. To alleviate this obstacle, Yang et al~\cite{yang2012approximation} proposed a linear system to approximate IC model, they verified in both theoretical and experimental aspects that IC model could be approximated as
\begin{equation}\label{eq:approxmation2IC}
  \textbf{f}^{IC}_{\overline{\mathcal{S}}}=(\textbf{I}-\textbf{T}_{\overline{\mathcal{S}\mathcal{S}}}')^{-1}\mathbf{t}_{\overline{\mathcal{S}}}
\end{equation}
when transmission matrix $\textbf{T}$ satisfies that $\textbf{T}\mathbf{e}<1$, where $\textbf{f}^{IC}_{\overline{\mathcal{S}}}$ is the vector of influences to the individuals not in $\mathcal{S}$ under IC model. Comparing Equation~\ref{eq:approxmation2IC} and Equation~\ref{eq:anotherdeduction}, we could find that, if we set $\Lambda=0$, $\textbf{f}^{IC}_{\overline{\mathcal{S}}}=\textbf{f}_{\overline{\mathcal{S}}}$. Actually, the approximation model in~\cite{yang2012approximation} is a specialization of linear circuit model. And linear circuit model could also approximate to IC model.

\subsubsection{Relationship with Aggarawal's Stochastic Model }
In 2011, Aggarawal et al~\cite{aggarwalflow} proposed a stochastic(ST) model to model the influence in a network which is as follows
\begin{equation}\label{eq:probability}
  f^{ST}_{\mathcal{S}\rightarrow i}=\left\{\begin{aligned}
    &1&\ \ \ \ &i\in \mathcal{S}&\\
    &1-\prod_{j=1}^n(1-t_{ji}f^{ST}_{\mathcal{S}\rightarrow j})&\ \ \ \ &i\not{\in}\mathcal{S}&
    \end{aligned}\right.
\end{equation}
where $t_{ji}$ is the transmission probability from $j$ to $i$ and $f^{ST}_{\mathcal{S}\rightarrow i}$ is the influence from $\mathcal{S}$ to $i$ under ST model. We can prove that
\begin{theorem}\label{th:stmodel}
If transmission matrix $\textbf{T}$ satisfies that $\textbf{T}\mathbf{e}\leq \mathbf{e}$, then for $i\not{\in}\mathcal{S}$,
\begin{equation}\label{eq:linearST}
f^{ST}_{\mathcal{S}\rightarrow i}=\frac{1}{1+\lambda_i}\sum_{i=1}^{n}{t_{ji} f^{ST}_{\mathcal{S}\rightarrow j}}\ \ \ \ \ \lambda_i\in [0,1)
\end{equation}
\end{theorem}
Theorem~\ref{th:stmodel} tells that ST model could also be approximated as a linear model and the damping coefficient on each individual should be ranged in $[0,1)$. Before proposing the proof of Theorem~\ref{th:stmodel}, we need to introduce a lemma first.
\begin{lemma}
  \label{lm:O_k(P)2}
  If denote $$O_k(P)=\sum_{i_1=1}^{n}\sum_{i_2=i_1+1}^{n}...\sum_{i_k=i_{k-1}+1}^{n}{p_{i_1}p_{i_2}...
p_{i_k}}$$ then
where $P=[p_1,p_2,...p_n]'$ and $\forall p_i\in [0,1]$, then
  \begin{equation}
    O_k(P)O_1(P)>2O_{k+1}(P)  \nonumber
  \end{equation}
  \begin{proof}
    Start from the left part of the inequality,
    \begin{eqnarray}
      &&O_k(P)O_1(P)=(\sum_{i_1=1}^{n}\cdot\cdot\cdot\sum_{i_k=i_{k-1}+1}^{n}{p_{i_1}\cdot\cdot\cdot p_{i_k}})(\sum_{i=1}^{n}p_{i})\nonumber\\
&&\ \ \ =\sum_{i_1=1}^{n}\cdot\cdot\cdot\sum_{i_k=i_{k-1}+1}^{n}\sum_{i_{k+1}=1}^{n}{p_{i_1}\cdot\cdot\cdot p_{i_k}p_{i_{k+1}}}\nonumber\\
&&\ \ \ =\sum_{i_1=1}^{n}\cdot\cdot\cdot\sum_{i_k=i_{k-1}+1}^{n}\sum_{i_{k+1}=1}^{i_k}{p_{i_1}\cdot\cdot\cdot p_{i_k}p_{i_{k+1}}}+O_{k+1}(P)\nonumber
    \end{eqnarray}
where
    \begin{eqnarray}
      &&\sum_{i_1=1}^{n}\cdot\cdot\cdot\sum_{i_k=i_{k-1}+1}^{n}\sum_{i_{k+1}=1}^{i_k}{p_{i_1}\cdot\cdot\cdot p_{i_k}p_{i_{k+1}}}=\nonumber\\
&&\ \ \ \ \ \ \sum_{i_1=1}^{n}\cdot\cdot\cdot\sum_{i_k=i_{k-1}+1}^{n}\sum_{i_{k+1}=1}^{i_{k-1}}{p_{i_1}\cdot\cdot\cdot p_{i_k}p_{i_{k+1}}}\nonumber\\
&&\ \ \ \ \ \ +\sum_{i_1=1}^{n}\cdot\cdot\cdot\sum_{i_k=i_{k-1}+1}^{n}\sum_{i_{k+1}=i_{k-1}+1}^{i_{k}}{p_{i_1}\cdot\cdot\cdot p_{i_k}p_{i_{k+1}}}\nonumber
    \end{eqnarray}
    Because of
    \begin{equation}
      \sum_{i_k=i_{k-1}+1}^{n}\sum_{i_{k+1}=i_{k-1}+1}^{i_{k}}{p_{i_k}p_{i_{k+1}}}=\sum_{i_k=i_{k-1}+1}^{n}\sum_{i_{k+1}=i_{k}+1}^{n}{p_{i_k}p_{i_{k+1}}},\nonumber
    \end{equation}
    we have
    \begin{eqnarray}
      &&\sum_{i_1=1}^{n}\cdot\cdot\cdot\sum_{i_k=i_{k-1}+1}^{n}\sum_{i_{k+1}=i_{k-1}+1}^{i_{k}}{p_{i_1}\cdot\cdot\cdot p_{i_k}p_{i_{k+1}}}\nonumber\\
&&\ \ \ \ \ \ \ =\sum_{i_1=1}^{n}\cdot\cdot\cdot\sum_{i_k=i_{k-1}+1}^{n}\sum_{i_{k+1}=i_{k}+1}^{n}{p_{i_1}\cdot\cdot\cdot p_{i_k}p_{i_{k+1}}}\nonumber\\
&&\ \ \ \ \ \ \ =O_{k+1}(P)\nonumber
    \end{eqnarray}

    Sum up the above analysis, there is
    $$
    O_k(P)O_1(P)>2O_{k+1}(P)
    $$
  \end{proof}
\end{lemma}
And if denote $p_j=t_{ji}f^{ST}_{\mathcal{S}\rightarrow j}$, Equation~\ref{eq:probability} for $i\not{\in}\mathcal{S}$ could be rewritten as
\begin{equation}
  \label{eq:probability2}
  f^{ST}_{\mathcal{S}\rightarrow i} = 1-\prod_{j=1}^n(1-p_j)=\sum_{k=1}^{n}(-1)^{k-1}O_k(P)
  \nonumber
\end{equation}
With this form and Lemma~\ref{lm:O_k(P)2}, we could prove Theorem~\ref{th:stmodel} now.

\textbf{The proof of Theorem~\ref{th:stmodel}}. Because $\textbf{T}$ satisfies that $\textbf{T}\mathbf{e}\leq\mathbf{e}$, there is $\sum_{j=1}^{n}{t_{ji}}\leq 1$, and $O_1(P)=\sum_{j=1}^{n}{t_{ji}f^{ST}_{\mathcal{S}\rightarrow j}}\leq\sum_{j=1}^{n}{t_{ji}}\leq 1$. With Lemma~\ref{lm:O_k(P)2}, we could get
\begin{equation}\label{eq:O_k(P)GEQ2}
O_k(P)>2O_{k+1}(P)
\end{equation}
For $f^{ST}_{\mathcal{S}\rightarrow i}=\sum_{k=1}^{n}(-1)^{k-1}O_k(P)=O_1(P)-O_2(P)+O_3(P)-...+(-1)^nO_n(P)$, it's easy to get
$$
f^{ST}_{\mathcal{S}\rightarrow i} \leq O_1(P)
$$
 and
$$
f^{ST}_{\mathcal{S}\rightarrow i} \geq O_1(P)-O_2(P) > O_1(P)-\frac{1}{2}O_1(P) = \frac{1}{2}O_1(P)
$$
In summary
$$
\frac{1}{2}O_1(P)<f^{ST}_{\mathcal{S}\rightarrow i}\leq O_1(P)
$$
which could be rewritten as
\begin{eqnarray}
f^{ST}_{\mathcal{S}\rightarrow i}&=&\eta_i O_1(P)=\eta_i\sum_{i=1}^{n}{t_{ji} f^{ST}_{\mathcal{S}\rightarrow j}}\ \ \ \ \ \eta_i\in (\frac{1}{2},1]\nonumber\\
&\stackrel{\frac{1}{1+\lambda_i}=\eta_i}{=}&\frac{1}{1+\lambda_i}\sum_{i=1}^{n}{t_{ji} f^{ST}_{\mathcal{S}\rightarrow j}}\ \ \ \ \ \ \ \ \ \ \  \ \lambda_i\in[0,1)\nonumber
\end{eqnarray}
It is proved.

\subsection{Rethinking Authority In the Perspective Of Influence}
\label{subsec:authority}
According to the dictionary, authority means the power of someone to influence the others. This interpretation gives a natural relation between influence and authority, that is, someone's authority is actually the total influence from her to the others. In the past years, the computation of authority for many things, such as web pages, facebook accounts, twitter accounts, has absorbed mountain of attentions due to its importance in the internet era. However, there is less work to discover the nature of authority. In this section, we will rethink the concept of authority in the perspective of influence and then propose a more accurate definition of it.

It's well accepted that pagerank algorithm and its variants, such as topic-sensitive pagerank, are the best methods to compute the authority of a node in a graph which could be internet, web network, twitter etc.
Based on~\cite{bianchini2005inside}, the general pagerank of nodes in a network could be formalized as follows. Denote $\mathbf{x}_t=[x^t_1,x^t_2,...x^t_n]'$ as the pagerank vector for all nodes on topic $t$, then
$$
\mathbf{x}_{t}=d\textbf{A}\mathbf{x}_t+\frac{(1-d)}{|\mathcal{S}_t|}\mathbf{e}_t
$$
where $d$ is a coefficient ranged in $(0,1)$,  $\textbf{A}=[a_{ij}]_{n*n}$ is a $n\times n$ matrix with $a_{ij}=\frac{w_{ji}}{OutDeg(j)}$ if there is an edge $(j,i)\in E$~\footnote{ $w_{ji}$ is the weight of edge $(j,i)$, usually it equals to 1} and 0 otherwise, $\mathcal{S}_t$ is the set of nodes which belong to topic $t$, and $\mathbf{e}_t=[e_1,e_2,...e_n]'$, where $e_i=1$ if node $i\in \mathcal{S}_t$ and 0 otherwise.  Notably, when $\mathcal{S}_t$ is a set with all nodes in the network (i.e. $\mathcal{S}_t = \mathcal{V}$), $\mathbf{x}_t$ will be the general pagerank vector.

The above equation could be solved as
\begin{eqnarray}\label{eq:tspr}
  \mathbf{x}_t&=&(I-d \textbf{A})^{-1}\frac{(1-d)}{|\mathcal{S}_t|}\mathbf{e}_t\nonumber\\
            &\stackrel{\frac{1}{d}=1+\lambda}{=}&(\textbf{I}+\lambda \textbf{I}-\textbf{A})^{-1}\lambda\frac{\mathbf{e_t}}{|\mathcal{S}_t|}\nonumber
\end{eqnarray}
where, for $d\in(0,1)$, $\lambda\in(0,+\infty)$. And $\sum_{i=1}^{n}{a_{ij}}=\sum_{i=1}^{n}{\frac{w_{ji}}{OutDeg(j)}}=1$, that is $\textbf{A}'\mathbf{e}=\mathbf{e}$, which means $\textbf{A}$ is a transmission matrix satisfying Assumption~\ref{as:theta}. Thus, if we view $\textbf{A}$ as $\textbf{T}$ and view $\lambda \textbf{I}$ as $\Lambda$, the matrix $(\textbf{I}+\lambda \textbf{I}-\textbf{A})^{-1}$ could be reviewed as the transpose of basis matrix $\textbf{P}$, that is
$$
\mathbf{x}_t = \frac{\lambda}{|\mathcal{S}_t|}\textbf{P}'\mathbf{e}_t
$$
which could be rewritten as, for $i=1,2,...n$
\begin{eqnarray}
x^t_i &=& \frac{\lambda}{|\mathcal{S}_t|}\sum_{j\in \mathcal{S}_t}{p_{ji}}\nonumber\\
&=&\frac{\lambda}{|\mathcal{S}_t|}p_{i\rightarrow\mathcal{S}_t}\nonumber\\
&\stackrel{Corollary~\ref{cor:fiVbound}}{=}&\frac{\lambda}{|\mathcal{S}_t|(1+\lambda)}b_{i\rightarrow \mathcal{S}_t}\nonumber\\
&\propto& b_{i\rightarrow \mathcal{S}_t}\label{eq:authorityessence}
\end{eqnarray}
From this equation, we could see that the node $i$'s pagerank value on topic $t$ is proportional to the upper bound of $f_{i\rightarrow \mathcal{S}_t}$. For the bound is consistently close to real influence, thus it could work well on the task of authority estimation. But in essence, the following assertion should be true.
\begin{assertion}
  The individual's authority on a group is essentially her total influences on each member of the group.
\end{assertion}
Based on it, we could propose a definition which may be more close to the nature of authority.
\begin{definition}\label{df:authority}
  The authority of $i$ on group $\mathcal{T}$ is equal to the sum of influences from $i$ to each member of $\mathcal{T}$, that is
  \begin{equation}\label{eq:authority}
    a_{i\rightarrow\mathcal{T}}=\sum_{j\in \mathcal{T}}f_{i\rightarrow j}
  \end{equation}
\end{definition}
With Equation~\ref{eq:fij},
$$
a_{i\rightarrow\mathcal{T}}=\sum_{j\in \mathcal{T}}\nu_{\{i\}\rightarrow i} p_{ji}=\nu_{\{i\},i} p_{i\rightarrow\mathcal{T}}\sim p_{i\rightarrow\mathcal{T}}
$$
This equation also could tell us why we call $p_{i\rightarrow\mathcal{T}}$ as the potential of $i$ influencing $\mathcal{T}$.

\section{An Application to Viral Marketing Problem}\label{sec:application}
Based on the discussion in Section~\ref{sec:theory}, viral marketing Problem, also called ass top-$K$ seeds selection problem or social influence maximization problem, which target at finding a small set of ``influential'' members of a network~(they are called as ``\textbf{seed}''s), could be formalized as the following optimization problem:
  \begin{equation}
  \mathcal{S}={\arg\max}_{\mathcal{S}\subset \mathcal{V}}f_{\mathcal{S}\rightarrow \mathcal{V}}\ \ \ \ subject\ to\ \ |\mathcal{S}|=K
  \nonumber
\end{equation}

In this problem, $f_{\mathcal{S}\rightarrow \mathcal{V}}$ is the influence from set $\mathcal{S}$ to set $\mathcal{V}$, in other words, is the expected number of individuals who will be influenced by members of $\mathcal{S}$ in the social network. This number is, conventionally, called as \textbf{influence spread} of $\mathcal{S}$ and denoted as $\sigma(\mathcal{S})$. $\sigma(\cdot)$ is a submodular function under IC model, that is
\begin{theorem}\label{th:submodular}
  For all the seeds set $\mathcal{S}\subseteq T\subseteq \mathcal{V}$ and any node $s$, it holds that
  \begin{equation}\label{eq:submodular}
    \sigma^{IC}(\mathcal{S}\cup \{s\})-\sigma^{IC}(\mathcal{S})\geq \sigma^{IC}(T\cup \{s\})-\sigma^{IC}(T)
  \end{equation}
  where $\sigma^{IC}(\mathcal{S})$ is the influence spread of $\mathcal{S}$ under IC model.
\end{theorem}

If denote $\Delta(\mathcal{S},s)=\sigma^{IC}(\mathcal{S}\cup \{s\})-\sigma^{IC}(\mathcal{S})$, as a corollary of Theorem~\ref{th:submodular}, there is
\begin{corollary}\label{cor:submodularseq}
Suppose $\mathcal{S}_0\subset \mathcal{S}_1\subset \mathcal{S}_2...\subset \mathcal{S}_K$ and $|\mathcal{S}_i|=i$, then
\begin{equation}
\Delta(\mathcal{S}_0,s) \geq \Delta(\mathcal{S}_1,s) \geq \Delta(\mathcal{S}_2,s)...\geq \Delta(\mathcal{S}_K,s)
\end{equation}
\end{corollary}
where $\Delta(\mathcal{S},s)$ denotes the marginal influence spread increment when adding $s$ into seed set $\mathcal{S}$.

\textbf{Proposed Algorithm.} As illustrated
in~\cite{kempe2003maximizing}, the optimization problem of top-$K$ seeds selection is NP-hard, and by exploiting the
submodular property of $\sigma(\mathcal{S})$, a greedy strategy guarantees to
obtain a solution that is within $(1-1/e)$ of the optimal result.
In a greedy framework, it always choose the individual who can produce the
maximal marginal increment on influence spread when adding her into $\mathcal{S}$. The
greedy algorithm starts with an empty set $\mathcal{S}_0=\emptyset$, and
iteratively, in each step $k$, adds $s_k$ who maximizes the
increment on influence spread into $\mathcal{S}_{k-1}$, that is
$$
s_k = {\arg\max}_{s\in \mathcal{V}\setminus \mathcal{S}_{k-1}}\Delta(\mathcal{S}_{k-1}, s)
$$
until the cardinality of seed set is $K$. Algorithm~\ref{algo:greedy} describes the greedy framework.

\begin{algorithm}
1. $\mathcal{S}=\emptyset$;\\
2. $s={\arg\max}_{s\in \mathcal{V}\setminus \mathcal{S}}{\Delta(\mathcal{S},s)}$;\\
3. $\mathcal{S}\cup=s$;\\
4. If $|\mathcal{S}|<K$, then go back to step2; else terminate.
 \caption{GreedyFramework}\label{algo:greedy}
\end{algorithm}

In the framework, step 2 is the most consuming step. Under IC model, to get $\Delta(\mathcal{S},s)$, the only available way is to run Monte-Carlo simulations of the model for a sufficiently many times~(e.g. 20,000). It is very inefficient.

Because linear circuit(LC) model could approximate to IC model~(see Section~\ref{subsec:relation2tranditionalmodels}), in this paper, we use $f^{LC}_{\mathcal{S}\rightarrow \mathcal{V}}$~(i.e., $f_{\mathcal{S}\rightarrow \mathcal{V}}$ discussed in Section~\ref{sec:theory}) to substitute for $\sigma(\mathcal{S})$, that is
\begin{equation}
  \label{eq:delta1}
  \Delta(\mathcal{S},s)\simeq\Delta^{I}(\mathcal{S},s)=f^{LC}_{\mathcal{S}\cup\{s\}\rightarrow \mathcal{V}}-f^{LC}_{\mathcal{S}\rightarrow \mathcal{V}}
\end{equation}
Based on the discussion in Section~\ref{subsec:anotherdeduction}, we know that $f^{LC}_{\mathcal{S}\rightarrow \mathcal{V}}$ could be computed in $O(|E|)$ time, thus $\Delta(\mathcal{S},s)$ could be computed in $O(|E|)$ under linear circuit model. 

Moreover, we could go on with this reduction work. Based on the discussion in Section~\ref{subsec:bound}, $b_{\mathcal{S}\rightarrow \mathcal{V}}$ is an estimation for $f^{LC}_{\mathcal{S}\rightarrow \mathcal{V}}$, then we could substitute $f^{LC}_{\mathcal{S}\rightarrow \mathcal{V}}$ by $b_{\mathcal{S}\rightarrow \mathcal{V}}$ further, that is
\begin{equation}
    \label{eq:delta2}
  \Delta(\mathcal{S},s)\simeq\Delta^{II}(\mathcal{S},s)=b_{\mathcal{S}\cup\{s\}\rightarrow \mathcal{V}}-b_{\mathcal{S}\rightarrow \mathcal{V}}
\end{equation}
With Equation~\ref{eq:bound}, this equation could be reformed as
\begin{equation}
    \label{eq:delta2actual}
  \Delta(\mathcal{S},s)\simeq(1+\lambda_s-\sum_{j\in \mathcal{S}}t_{js})p_{s\rightarrow\mathcal{V}}-\sum_{j\in \mathcal{S}}t_{sj}p_{j\rightarrow\mathcal{V}}
\end{equation}
Since $\textbf{p}_\mathcal{V}=[p_{1\rightarrow\mathcal{V}},p_{2\rightarrow\mathcal{V}},...,p_{n\rightarrow\mathcal{V}}]'$ is a quantity that could be computed in advance~(see Section~\ref{subsec:bound}), thus, for any $\mathcal{S}$ and any $s$, the computation of $\Delta(\mathcal{S},s)$ in Equation~\ref{eq:delta2actual} only spends $O(|\mathcal{S}|)$ time.

Along this reduction way, we could get more profit. Based on Corollary~\ref{cor:fiVbound}, there is
$
\Delta^{I}(\mathcal{S}_0,s)=f^{LC}_{s\rightarrow \mathcal{V}}\leq (1+\lambda_s)p_{s\rightarrow\mathcal{V}}
$
and
$
\Delta^{II}(\mathcal{S}_0,s)=b_{s\rightarrow \mathcal{V}}= (1+\lambda_s)p_{s\rightarrow\mathcal{V}}
$, thus if substituting $\Delta(\mathcal{S},s)$ by $\Delta^{I}(\mathcal{S},s)$ or $\Delta^{II}(\mathcal{S},s)$, Corollary~\ref{cor:submodularseq} could be reformed as
$$
(1+\lambda_s)p_{s\rightarrow\mathcal{V}}\leq \Delta(\mathcal{S}_0,s)\leq \Delta(\mathcal{S}_1,s)...\leq\Delta(\mathcal{S}_K,s)
$$
which means that the marginal influence increment of individual $s$ can not be larger than $(1+\lambda_s)p_{s\rightarrow\mathcal{V}}$ and her marginal increment in previous iterations. Thus, in each iteration of Algorithm~\ref{algo:greedy}, when we go into step 2, we have at least one upper bound for estimating $\Delta(\mathcal{S},s)$, that is either $(1+\lambda_s)p_{s\rightarrow\mathcal{V}}$ or $s$'s increment in last iteration. Upper bound, as an estimation for real value, can help us to reduce chunk of vain computations. For example, if the upper bound of an individual's influence increment is not large enough~(comparably), it is impossible that she will make $\Delta(\mathcal{S},s)$ maximal, thus we could just skip the individual. Actually, this property of upper bound could help us to skip many individuals with small influence and sharply reduce the total computation time. The complete procedures for viral marketing problem is illustrated in Algorithm~\ref{algo:Circuit}.
\begin{algorithm} [th]
\SetKwInOut{Input}{input}
\SetKwInOut{Output}{output}
\Input{$\mathcal{G}(\mathcal{V},\mathcal{E}),K, \Lambda, \textbf{T}$\\}
\Output{$\mathcal{S}$\\}
$\mathcal{S}=\emptyset$;\\
Compute authority vector $\textbf{p}_{\mathcal{V}}=[p_{1\rightarrow\mathcal{V}}, p_{2\rightarrow\mathcal{V}},...p_{n\rightarrow\mathcal{V}}]'$~(see it in Section~\ref{subsec:bound});\\
\For{each vertex $s$ in $\mathcal{G}$}
{
    $\Delta_s=(1+\lambda_s)p_{s\rightarrow\mathcal{V}}$;
}
\While{$|\mathcal{S}|<K$}
{
    re-arrange the order of node to make $\Delta_s\geq \Delta_{s+1}$;\\
    $\Delta_{max}=0$;\\
    \For{$s=1$ to $n-|\mathcal{S}|$}
    {
        \If{$\Delta_s>\Delta_{max} $}
        {
            $\Delta_s$ = GetDeltaI$(\mathcal{S}, s, f_{\mathcal{S}\rightarrow \mathcal{V}}) \Lambda, \textbf{T}$\\
            //or $\Delta_s$= GetDeltaII$(\mathcal{S}, s, \Lambda, \textbf{T}, \textbf{p}_{\mathcal{V}})$;\\
            \If{$\Delta_s>\Delta_{max} $}
            {
                $\Delta_{max} =\Delta_s$;\\
                $s_{max}=s$;\\
            }
        }
        \Else
        {
            break;\\
        }
    }
    $\mathcal{S}= \mathcal{S}\cup \{s\}$;\\
    $f_{\mathcal{S}\rightarrow \mathcal{V}} = f_{\mathcal{S}\rightarrow \mathcal{V}}+\Delta_{max}$;\\
    $\Delta_s = 0$;\\
}
return $\mathcal{S}$;\\
\caption{LinearCircuitMethod($\mathcal{G}$, $K$, $\Lambda$, $\textbf{T}$)}\label{algo:Circuit}
\end{algorithm}

\begin{function}
\SetKwInOut{Input}{input} \SetKwInOut{Output}{output}
\Input{$\mathcal{S}, s, f_{\mathcal{S}\rightarrow \mathcal{V}}, \Lambda, \textbf{T}$}
\Output{$\Delta_s$\\}
If $\textbf{P}_{\cdot s}$ has never been computed, compute it first~(see it in Section~\ref{subsec:computation});\\
$\mathcal{S}'=\mathcal{S}\cup\{s\}$;\\
Compute $\nuup_{\mathcal{S}'\mathcal{S}'}=\textbf{P}_{\mathcal{S}'\mathcal{S}'}^{-1}\mathbf{e}$~(Theorem~\ref{th:solution});\\
$\textbf{f}_{\mathcal{S}'}=\mathbf{0}$, $f_{\mathcal{S}'\rightarrow \mathcal{V}}=0$;\\
\For{int $j\in \mathcal{S}'$}
{
$\textbf{f}_{\mathcal{S}'}\ +=\ \nu_j \textbf{P}_{\cdot j}$;//Equation~\ref{eq:FSlinear}\\
}
\For{each $j\in \mathcal{V}$}
{
$f_{\mathcal{S}'\rightarrow \mathcal{V}}\ +=\ f_{\mathcal{S}'\rightarrow j}$;//Axiom~\ref{ax:1}\\
}
return $f_{\mathcal{S}'\rightarrow \mathcal{V}}-f_{\mathcal{S}\rightarrow \mathcal{V}}$;\\
 \caption{GetDeltaI($\mathcal{S}$, $s$, $f_{\mathcal{S}\rightarrow \mathcal{V}}$, $\Lambda$, $\textbf{T}$)}\label{func:getdeltaI}
\end{function}

\begin{function}
\SetKwInOut{Input}{input} \SetKwInOut{Output}{output}
\Input{$\mathcal{S}, s, \Lambda, \textbf{T},\textbf{p}_{\mathcal{V}}$}
\Output{$\Delta_s$\\}
$\Delta_s=(1+\lambda_s)p_{s\rightarrow\mathcal{V}}$;\\
\For{each $j\in \mathcal{S}$}
{
$\Delta_s=\Delta_s-t_{js}p_{s\rightarrow \mathcal{V}} - t_{sj}p_{j\rightarrow \mathcal{V}}$;
}
return $\Delta_s$;\\
 \caption{GetDeltaII($\mathcal{S}$, $s$, $\Lambda$, $\textbf{T}$, $\textbf{p}_{\mathcal{V}}$)}\label{func:getdeltaII}
\end{function}

In Algorithm~\ref{algo:Circuit}, we use $\Delta_s$ to store the upper bound of $\Delta(\mathcal{S},s)$ and use $\Delta_{max}$ and $s_{max}$ to store the maximal $\Delta(\mathcal{S},s)$ and its corresponding $s$. The algorithm starts with an empty set $\mathcal{S}=\emptyset$, at this moment $\Delta_s=(1+\lambda_s)p_{s\rightarrow\mathcal{V}}$ and in each iteration, it adds the $s$ with the maximal $\Delta(\mathcal{S},s)$ into $\mathcal{S}$ until the carnality of $\mathcal{S}$ is equal to $K$. Specifically, in each iteration, we first re-arrange the index of individual to make $\Delta_s\geq \Delta_{s+1}$ which can help us to aim at those individuals with big $\Delta$ at the beginning and reduce those vain computation spending on nobody; then, for each individual $i$, we compare her upper bound $\Delta_s$ with $\Delta_{max}$. 1) If it is larger, then $i$ maybe a better one, then we need to compute its real increment; if her real increment is still larger than $\Delta_{max}$, then this one is truly a better one, then we store her index by variable $s_{max}$ and store her real increment into $\Delta_{max}$. 2) If it is smaller, then $s$ and all of her successors cannot be better than the current $s_{max}$ for $\Delta_{max} \geq\Delta_s\geq\Delta_{s+1}$; then, we can break out of the iteration. When out of an iteration, the index of best supplemental individual has been stored in variable $s_{max}$, we just need to add it into $\mathcal{S}$, at the same time, we should add the real increment by $s_{max}$ into $f_{\mathcal{S}\rightarrow \mathcal{V}}$ also. At last, set $\Delta_{s_{max}}$ to be 0, then, individual $s_{max}$ could not be accessed again.
According to the way how to compute $\Delta(\mathcal{S},s)$, we call the Algorithm with Function~\ref{func:getdeltaI} as \textbf{Circuit\_Complete(CC)} method, and call the one with Function~\ref{func:getdeltaII} as \textbf{Circuit\_Fast(CF)} method.

\section{Experiment Part}\label{sec:exp}
In this section, we will do the following experiments: a) Comparing linear circuit model with IC model and ST model to verify the relationship among them; b) Demonstrate the effectiveness of upper bound $b_{\mathcal{S}\rightarrow \mathcal{T}}$ to estimate $f_{\mathcal{S}\rightarrow \mathcal{T}}$; c) evaluate the performances of \textbf{Circuit\_Complete} and \textbf{Circuit\_Fast} algorithm and compare them with the state-of-the-art algorithms on real-world social networks.

\begin{figure*}
  \begin{center}
    \subfigure[ \vspace{-1cm}  Polblogs.
    \quad]{\label{fig:wikiEffi}\includegraphics[scale=0.28]{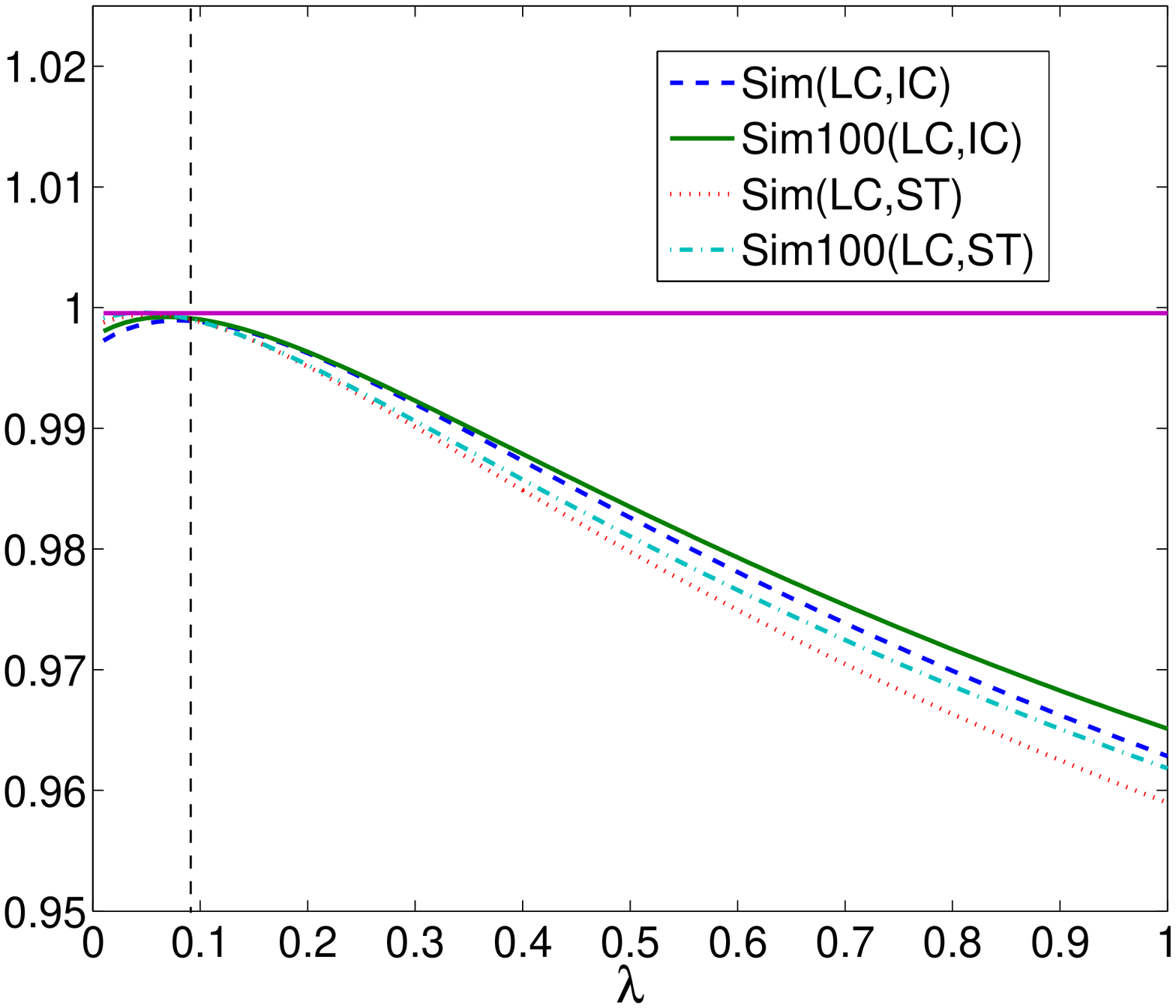}}  
    \subfigure[ \vspace{-1cm}  Wiki-Vote.
    \quad]{\label{fig:PHYEffi}\includegraphics[scale=0.28]{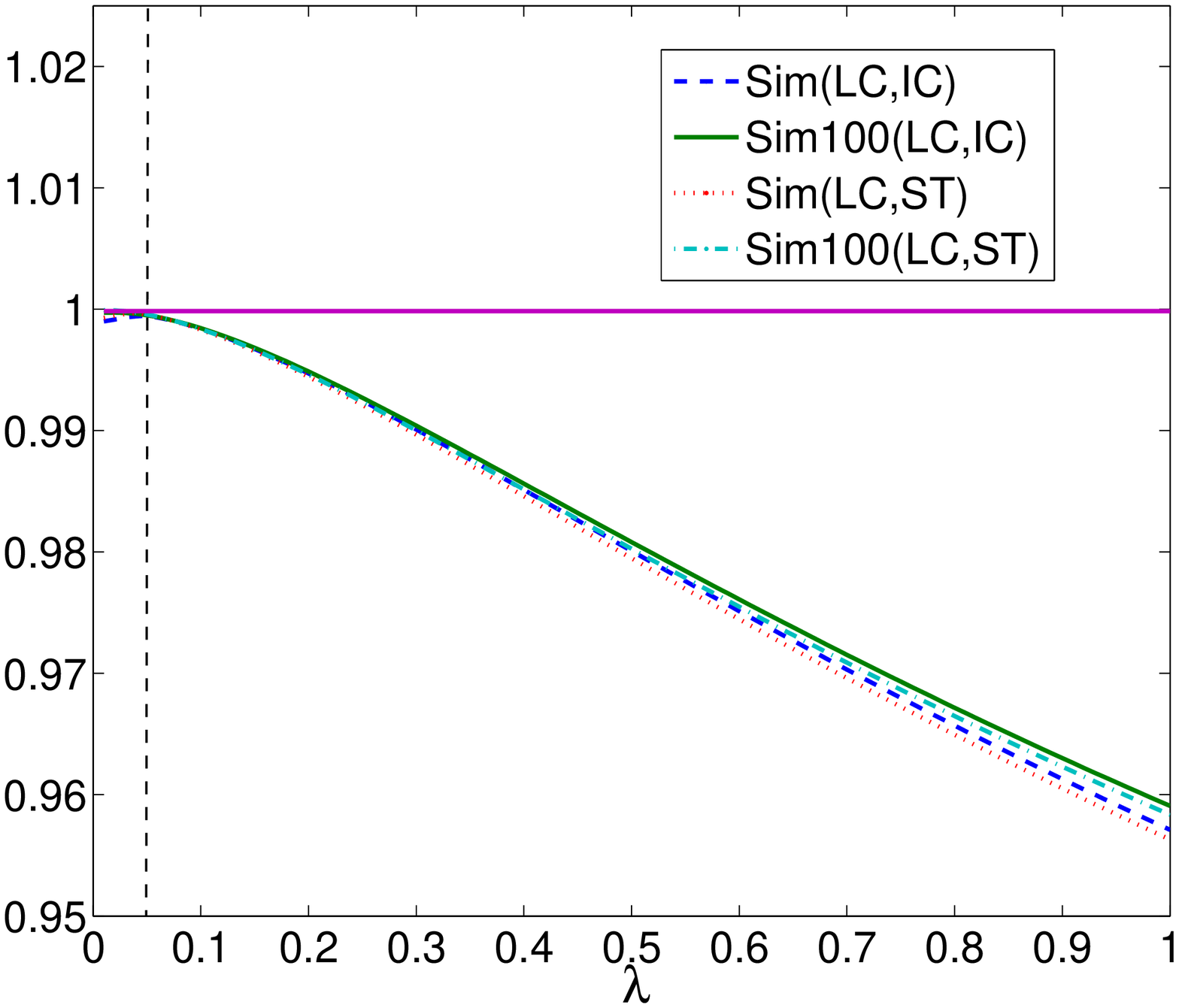}}  
    \subfigure[ \vspace{-1cm}  ca-HepPh.
    \quad]{\label{fig:PHYEffi}\includegraphics[scale=0.28]{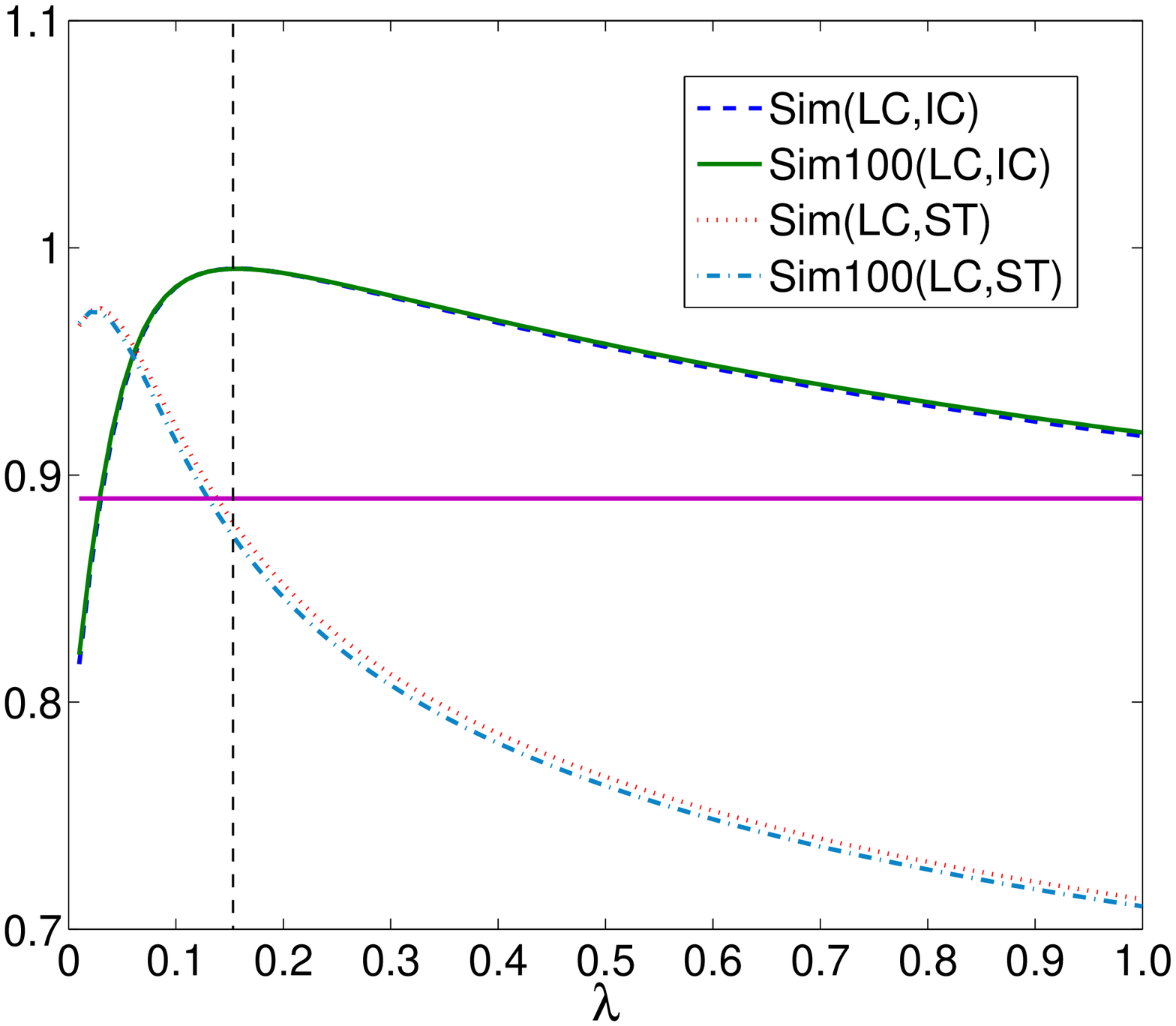}}   
  \end{center}
\vspace{-0.4cm}
  \caption{The cosine similarity between LC model and the other two models on Polblogs, Wiki-Vote, ca-HepPh respectively. In the three subfigures, the black vertical dash line is a mark of the optimal value of similarity; and the purple horizontal line is the level of similarity between IC model and ST model. $Sim100(A,B)$ is the similarity between A model and B model on 100 randomly selected sets.\label{fig:modelcomp}}
\vspace{-0.6cm}
\end{figure*}

\subsection{Date Sets}
The first data, denoted as \textbf{Polblogs}, is a directed network of hyperlinks between weblogs on US politics, recorded in 2005 by Adamic and Glance~\cite{adamic2005political}. There are 1,499 nodes and 19,090 edges in this network.

The second data, denoted as \textbf{Wiki-Vote}, is a Wikipedia voting network in which nodes represent wikipedia users and a directed edge from node $i$ to node $j$ represents that user $i$ voted on user $j$, the network contains all
the Wikipedia voting data from the inception of Wikipedia till January 2008~\footnote{http://snap.stanford.edu/data/wiki-Vote.html}. This directed network contains 7,115 nodes and 103,689 edges.

The third one, denoted as \textbf{ca-HepPh}, is a collaboration network which is from the e-print arXiv which covers scientific collaborations between authors whose papers have been submitted to \emph{High Energy Physics - Phenomenology category}~\footnote{http://snap.stanford.edu/data/ca-HepPh.html}. This undirected network contains 12,008 nodes and 654,188 edges.

The fourth one, denoted as \textbf{DBLP}, is an even larger collaboration
network, the DBLP Computer Science Bibliography Database, which is the same as in~\cite{chen2010scalable}. This undirected network contains 655,000 nodes and 1,967,265 edges.

The fifth one, denoted as \textbf{web-NotreDame}, is an webpage link network where nodes represent pages from University of Notre Dame (domain nd.edu) and directed edges represent hyperlinks between them. The data was collected in 1999 by Albert, Jeong and Barabasi~\footnote{http://snap.stanford.edu/data/web-NotreDame.html}. This directed network contains 325,729 nodes and 1,497,134 edges.

The sixth one, denoted as \textbf{LiveJournal}, is a friendship network crawled from LiveJournal~\footnote{http://www.livejournal.com} on July, 2010~\cite{Zafarani+Liu:2009}. This is a large-scale network, containing 2,238,731 nodes and 14,608,137 edges.

We chose these networks since they can cover a variety of networks
with sizes ranging from 103K edges to 14M edges and include four
directed networks and two undirected networks.

\subsection{Model Similarity}\label{subsec:modelcomp}

In Section~\ref{subsec:relation2tranditionalmodels}, we proved that linear circuit(LC) model is closely related to independent cascade model(IC) and stochastic(ST) model. In this section, we will verify their relationship by experimental results.  Suppose $\textbf{f}_\mathcal{S}^{LC}$, $\textbf{f}_{\mathcal{S}}^{IC}$, and $\textbf{f}_{\mathcal{S}}^{ST}$ are the influence vector of seed set $\mathcal{S}$ under LC model, IC model and ST model respectively. If LC model is closely related to the other two models, $\textbf{f}_\mathcal{S}^{LC}$ must be similar with $\textbf{f}_{\mathcal{S}}^{IC}$ and $\textbf{f}_{\mathcal{S}}^{ST}$ for any set $\mathcal{S}$, and vice versa. In this paper ,we use Cosine similarity as the metric to measure the similarity among $\textbf{f}_\mathcal{S}^{LC}$ and $\textbf{f}_{\mathcal{S}}^{IC}$, $\textbf{f}_{\mathcal{S}}^{ST}$. That is,
\begin{equation}
  Sim(\textbf{f}_\mathcal{S}^{A},\textbf{f}_\mathcal{S}^{B})= Cos(\textbf{f}_\mathcal{S}^{A},\textbf{f}_\mathcal{S}^{B})
\end{equation}
where, $A,B$ are indicators for model and $Cos$ is Cosine function. Along this line, we propose a formula to define the similarity between models,
\begin{equation}
  Sim(A,B)=\frac{\sum_{\mathcal{S}\subset \mathcal{V}}Sim(\textbf{f}_\mathcal{S}^{A}, \textbf{f}_\mathcal{S}^{B})}{\sum_{\mathcal{S}\subset \mathcal{V}}}\label{eq:modelsim1}
\end{equation}
This equation is very exhaustive to be computed for there are $2^{|\mathcal{V}|}$ choices for $\mathcal{S}$. However, practically, we could randomly selected a certain number of sets as representation of the all to get an approximation of $Sim(A,B)$. 

On three datasets, \textbf{polblogs, Wiki-Vote, ca-HepPh}, we compute the similarities between models under the following settings:
\begin{enumerate}
  \item randomly select 50,000 sets as representation of the all sets;
  \item set $\Lambda=\lambda \textbf{I}$~\footnote{It means that the damping coefficients of all individuals are identical, which is for a global model but not a personalized model.} where $\lambda$ ranges in $[0,1)$~(see into Section~\ref{subsec:modelcomp}), starts from 0.01 and steps by 0.01;
  \item set $\textbf{T}=\textbf{D}^{-1}\textbf{W}'$~\footnote{In this setting, $\textbf{T}\textbf{e}=\textbf{D}^{-1}\textbf{W}'\textbf{e}=\textbf{I}$ which satisfies Assumption~\ref{as:theta}, and $t_{ij}=\frac{w_{ji}}{\sum_k{w_{jk}}}$.}, where $\textbf{W}$ is the trust weight matrix of $\mathcal{G}$ and $\textbf{D}=diag(\textbf{W}'\textbf{e})$.
\end{enumerate}
\begin{figure*}
  \begin{center}
    \subfigure[ \vspace{-1cm}  Polblogs.
    \quad]{\label{fig:wikiEffi}\includegraphics[scale=0.259]{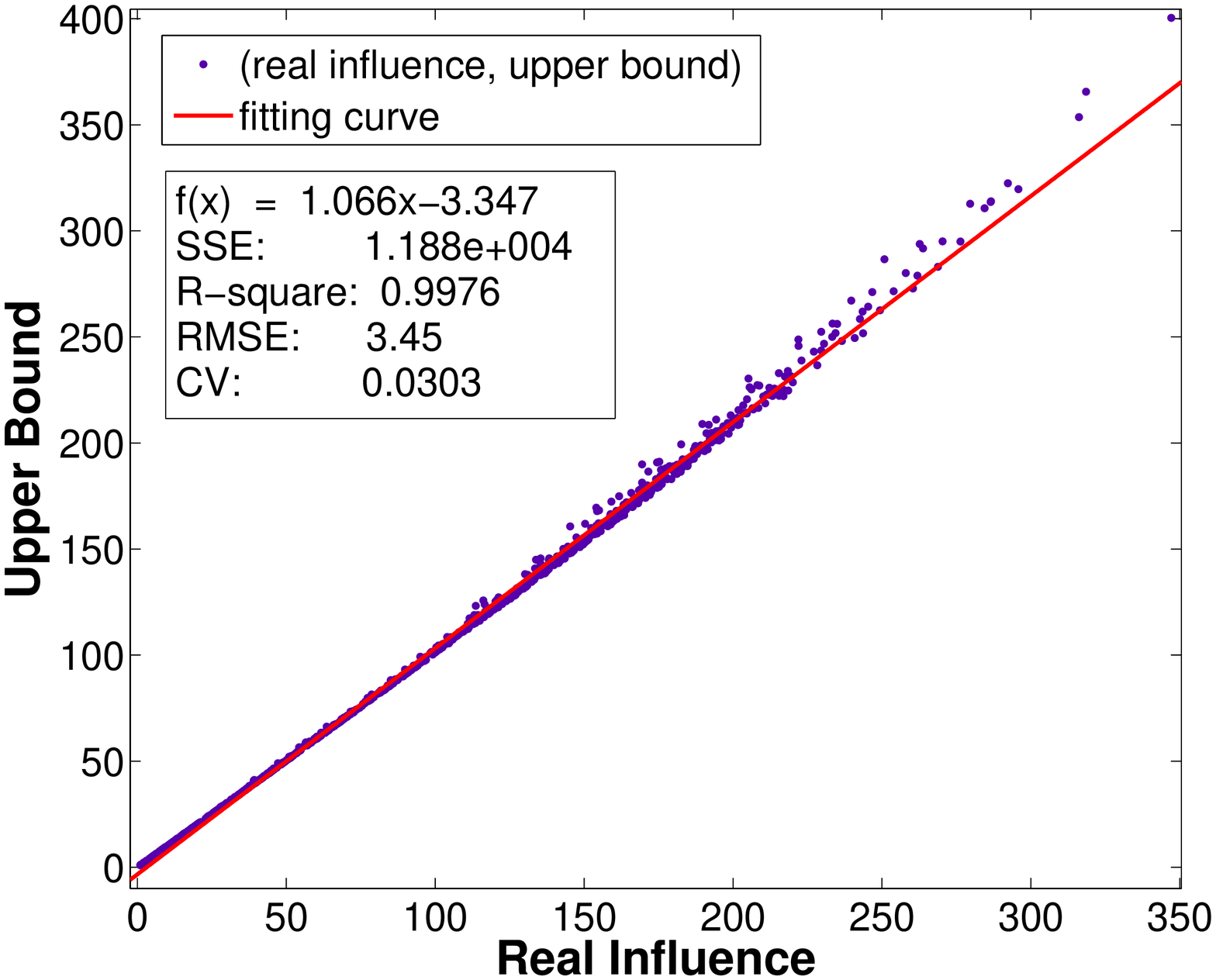}} 
    \subfigure[ \vspace{-1cm}  Wiki-Vote.
    \quad]{\label{fig:PHYEffi}\includegraphics[scale=0.259]{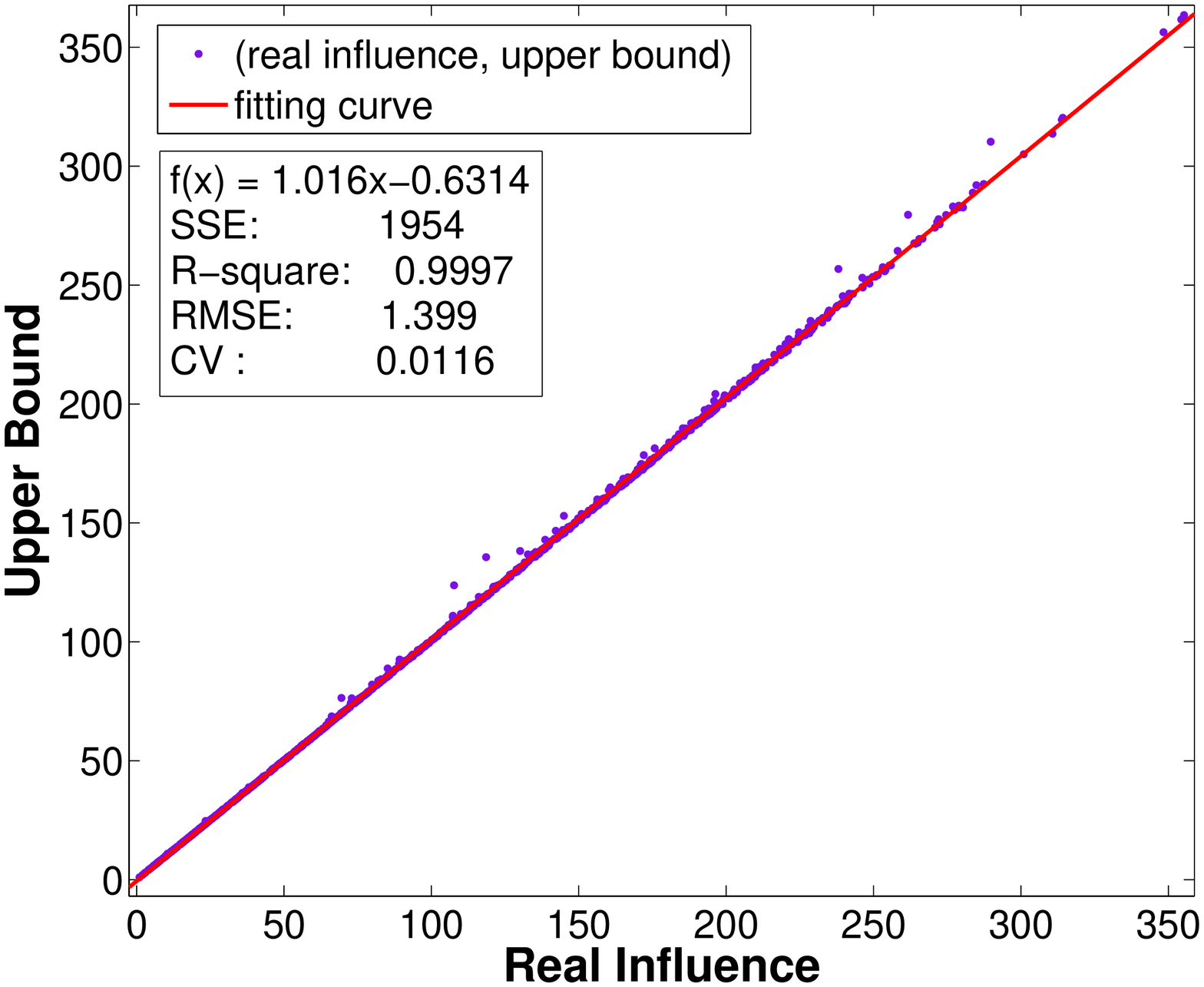}} 
    \subfigure[ \vspace{-1cm}  ca-HepPh.
    \quad]{\label{fig:PHYEffi}\includegraphics[scale=0.252]{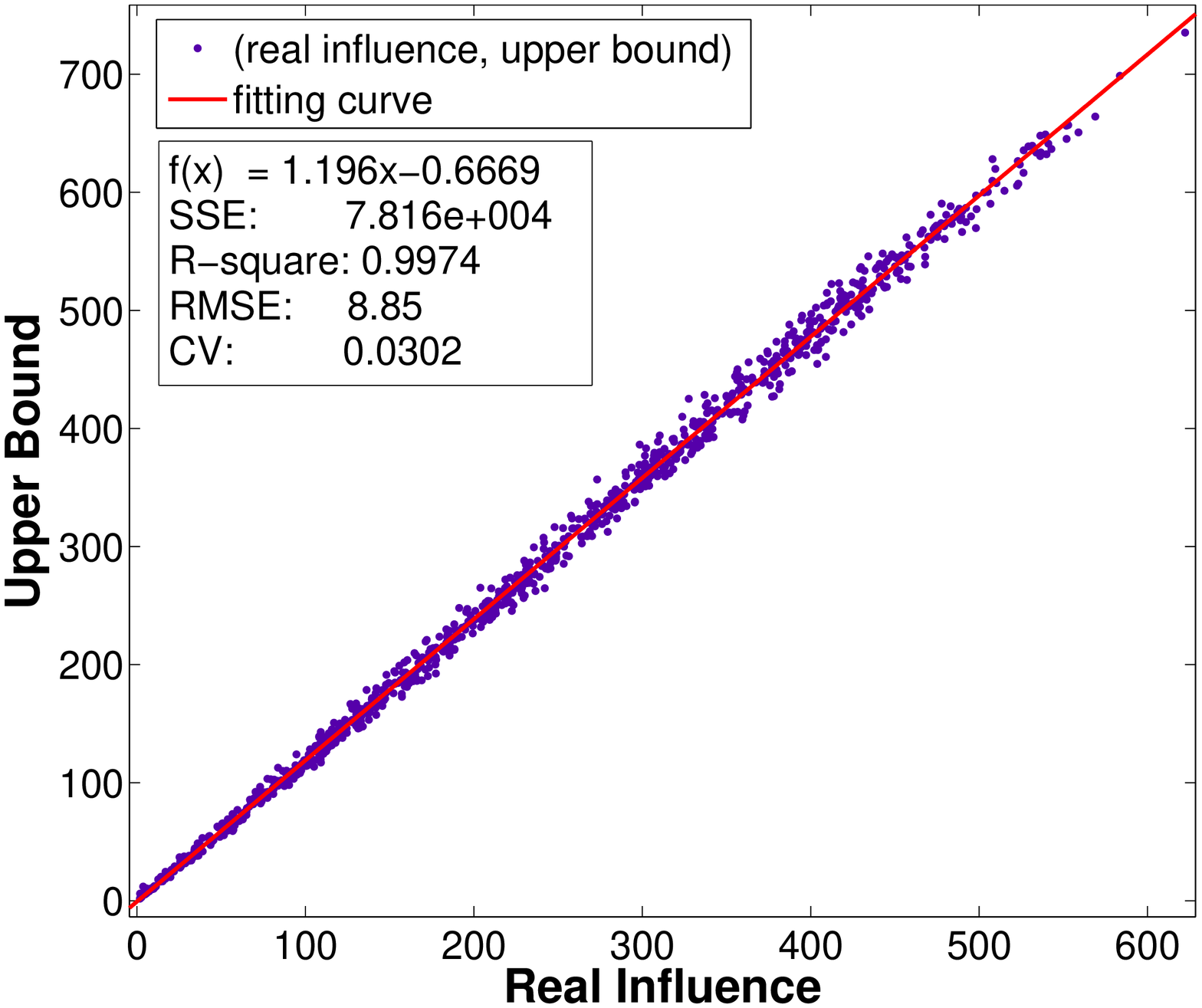}} 
  \end{center}
\vspace{-0.4cm}
  \caption{The plots of ($f_{\mathcal{S}\rightarrow \mathcal{T}}$,$b_{\mathcal{S}\rightarrow\mathcal{T}}$) and their fitting curve on Polblogs, Wiki-Vote, ca-HepPh respectively. \label{fig:fitting}}
\vspace{-0.6cm}
\end{figure*}
The experimental results is shown in Figure~\ref{fig:modelcomp}. We could get the following observations:
\begin{itemize}
  \item On each data sets, the similarity between LC model and IC model could reach a high level~(even larger than 0.99); the similarity between ST model and the other two models is not very stable on different data sets, e.g., on \textbf{ca-HepPh} the similarity between ST and IC model is only 0.88;
  \item The similarity curves between LC model and IC model all increased firstly and then decreased, and reached their peaks at a certain $\lambda$ ranges in [0.05,0.15];
  \item When $\lambda$ ranges in [0.10, 0.30], the similarity between LC model and IC model keeps in a high level~(always larger than 0.97) on every data sets; 
  \item The curve of $Sim(A,B)$ and $Sim100(A,B)$ is very close. The similarity curve computed on randomly selected 50,000 sets makes little difference with the similarity on 100 sets.
\end{itemize}
These observations tell us that LC model might be a proper approximation to IC model; if we want to use LC model to approximate to IC model, we'd better assign a value ranging in [0.10,0.30] to $\lambda$; if we want to work out whether or not a $\lambda$ value is good to approximate to IC model or ST model, we just need to test it by computing the model similarity under a few number of sample sets. Based on the above observations, we also could explain why it is proper to use $f_{\mathcal{S}\rightarrow \mathcal{V}}$ to replace $f^{IC}_{\mathcal{S}\rightarrow \mathcal{V}}$ in Section~\ref{sec:application}.

\subsection{The Comparison Between $f_{\mathcal{S}\rightarrow \mathcal{T}}$ and $b_{\mathcal{S}\rightarrow \mathcal{T}}$}\label{subsec:expbound}
In Section~\ref{subsec:bound}, we proved that $b_{\mathcal{S}\rightarrow \mathcal{T}}=\sum_{j\in \mathcal{S}}((1+\lambda_j)-\sum_{k\in \mathcal{S}}t_{kj})p_{j\rightarrow\mathcal{T}}$ is an upper bound of $f_{\mathcal{S}\rightarrow \mathcal{T}}$. In this section, we will explore the relationship between them in the aspect of experimental investigation. On the three data sets, \textbf{polblogs, Wiki-Vote, ca-HepPh}, we compute 1,000 pairs of $(f_{\mathcal{S}\rightarrow \mathcal{T}},b_{\mathcal{S}\rightarrow \mathcal{T}})$ respectively and plot them into three coordinates. The three coordinates are shown in Figure~\ref{fig:fitting} where the read lines are the optimal linear curve fitting for those plots. From Figure~\ref{fig:fitting}, we could observe that
\begin{itemize}
  \item The upper bound $b_{\mathcal{S}\rightarrow \mathcal{T}}$  is almost linearly correlated to influence $f_{\mathcal{S}\rightarrow \mathcal{T}}$, when fitting a linear line to the dots, the coefficient of variants are only 0.0303, 0.0116, 0.0302 respectively;
  \item The upper bound $b_{\mathcal{S}\rightarrow \mathcal{T}}$ is consistently close to influence $f_{\mathcal{S}\rightarrow \mathcal{T}}$, the gradient of the three fitting lines are 1.066, 1.016 and 1.196, which means that in average $b_{\mathcal{S}\rightarrow \mathcal{T}}$ only exceeds $f_{\mathcal{S}\rightarrow \mathcal{T}}$ 6.6\%, 1.6\%, and 19.6\% respectively;
\end{itemize}
Because $b_{\mathcal{S}\rightarrow \mathcal{T}}$ is consistently close to $f_{\mathcal{S}\rightarrow \mathcal{T}}$, it is feasible to substitute $b_{\mathcal{S}\rightarrow \mathcal{T}}$ for $f_{\mathcal{S}\rightarrow \mathcal{T}}$. For the computation cost of $b_{\mathcal{S}\rightarrow \mathcal{T}}$ is much less than $f_{\mathcal{S}\rightarrow \mathcal{T}}$, thus when we make this substitution in practice, the computation cost of real application is probably to be sharply reduced. By the way, the experimental settings of this part is in some way along with the settings of above section: \begin{itemize}
  \item $\Lambda=\lambda\textbf{I}$ and $\lambda=0.2$;
  \item $\mathcal{T}$ is set to be $\mathcal{V}$;
  \item $\textbf{T}=\textbf{D}^{-1}\textbf{W}'$.
\end{itemize}

\subsection{Viral Marketing Campaign Design~(or Top-K Seeds Selection)}
In this section, we will use \textbf{Circuit\_Complete~(CC)} and \textbf{Circuit\_Fast~(CF)} to face the challenge of viral marketing problem and compare them with the state-of-the art algorithms to verify their effectiveness and efficiency.

\textbf{Benchmark Algorithms.} The benchmark algorithms for viral marketing problem are as follows. First, \textbf{Circuit\_Independent~(CI)} is the Algorithm proposed in~\cite{biao2012linear}. {\textbf{CELF}} is the original greedy algorithm with the CELF optimization of \cite{leskovec2007cost}, where the times of Monte-Carlo simulations is set to be 20000. \textbf{PMIA} is the algorithm proposed in~\cite{chen2010scalable}. We used the source code provided by the authors, and set the parameters to the ones produce the best results~\footnote{Based on the source code from its author, the parameter would be selected from \{1/10,1/20,1/40,1/80,1/160,1/320,1/1280\}}. In the \textbf{PageRank~(PR)} algorithm~\cite{page1999pagerank}, we selected top-$K$ nodes with the highest pagerank value. \textbf{DegreeDiscountIC~(DIC)}~\cite{chen2009efficient} measures the degree discount heuristic with a propagation probability of $p = 0.01$, which is the same as used in \cite{chen2009efficient}. Finally, the \textbf{Degree~(Deg)} method captures the top-$K$ nodes with the highest degree. Among these algorithms, Degree, DegreeDiscountIC and pageRank are
widely used for baselines. To the best of our knowledge, CELF and
PMIA are two of the best existing algorithms in terms of solving
the viral marketing problem~(concerning the tradeoff between
effectiveness and efficiency).

\textbf{Measurement.} The effectiveness of the algorithms for the viral marketing problem is justified by the estimated number of individuals that will be influenced by the chosen seed set of each algorithm, i.e., influence spread $\sigma(\mathcal{S})$. To estimate the influence spread, for each seed set, we run the Monte-Carlo simulation under independent cascade model~\footnote{In detail, under the IC model, the node in the seed set propagates its influence through the following operations. Let us view the node in the seed set $\mathcal{S}$ as the node influenced at time $t=0$,  if node $i$ is influenced at time $t$, then it will influence its not-yet-influenced neighbor node $j$ at time $t+1$ (and only time $t+1$) with transmission probability $t_{ij}$. In this paper, as long as the transmission probabilities on edges satisfy the confinement of Assumption~\ref{as:theta}, our method will handle its corresponding influence maximization problem. Due to the limited space, in this paper, we set the transmission probability $t_{ij}$ as equal to $\frac{c_{ij}}{d_j}$ which is widely adopted in the previous studies and its corresponding model is called as Weighted Cascade (WC) Model.} 20000 times to find how many individuals can be
influenced, and then use these influence spreads to
compare the effectiveness of these algorithms. 

{\bf Experimental Platform}. The experiments were performed
on a server with 2.0GHz Quad-Core Intel Xeon E5410 and 8G memory.

\subsubsection{A performance comparison}
\begin{figure*}
  \begin{center}
    \subfigure[  Polblogs.
    \quad]{\label{fig:wiki}\includegraphics[scale=0.25]{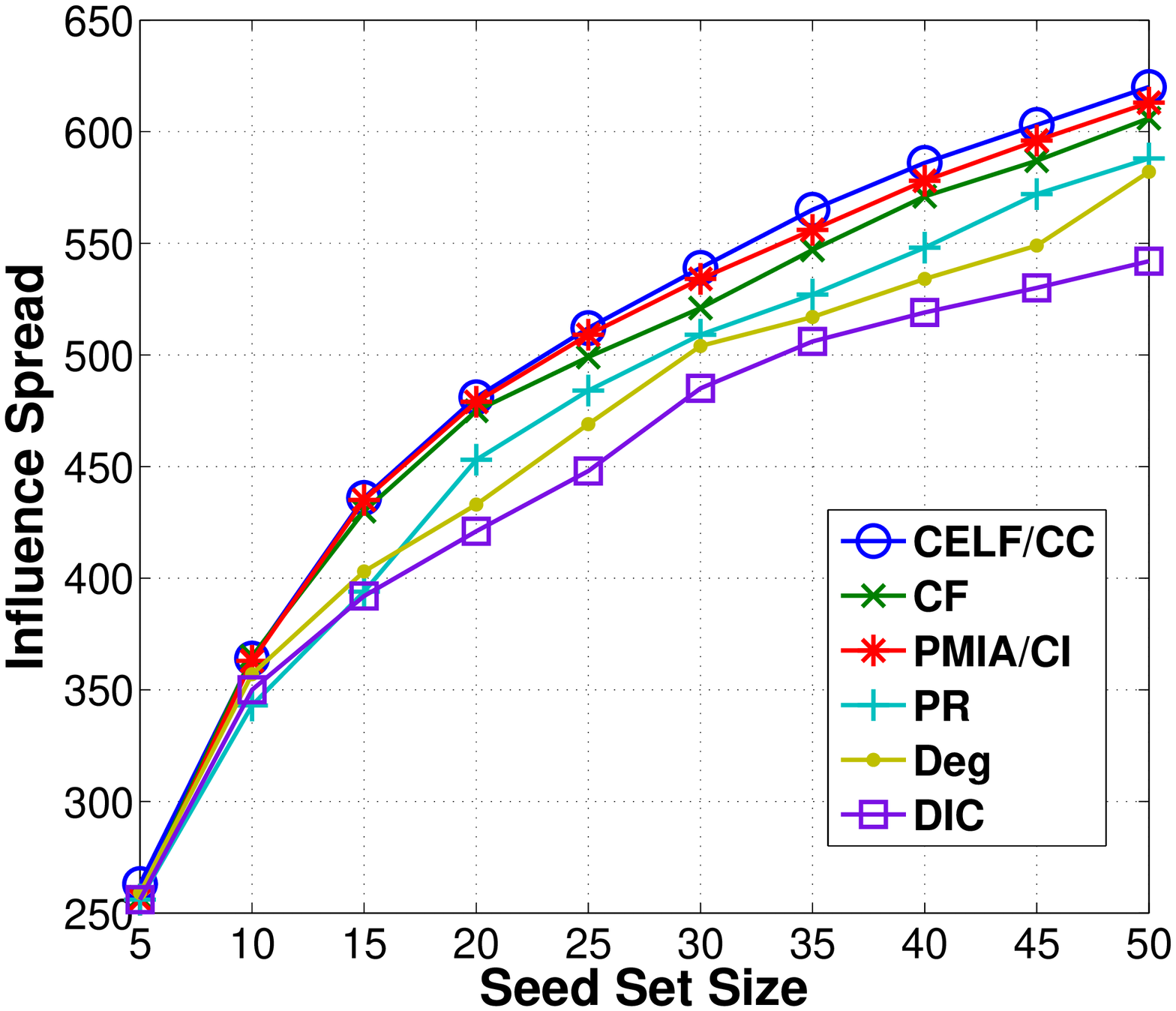}}  
    \subfigure[  Wiki-Vote.
    \quad]{\label{fig:phy}\includegraphics[scale=0.25]{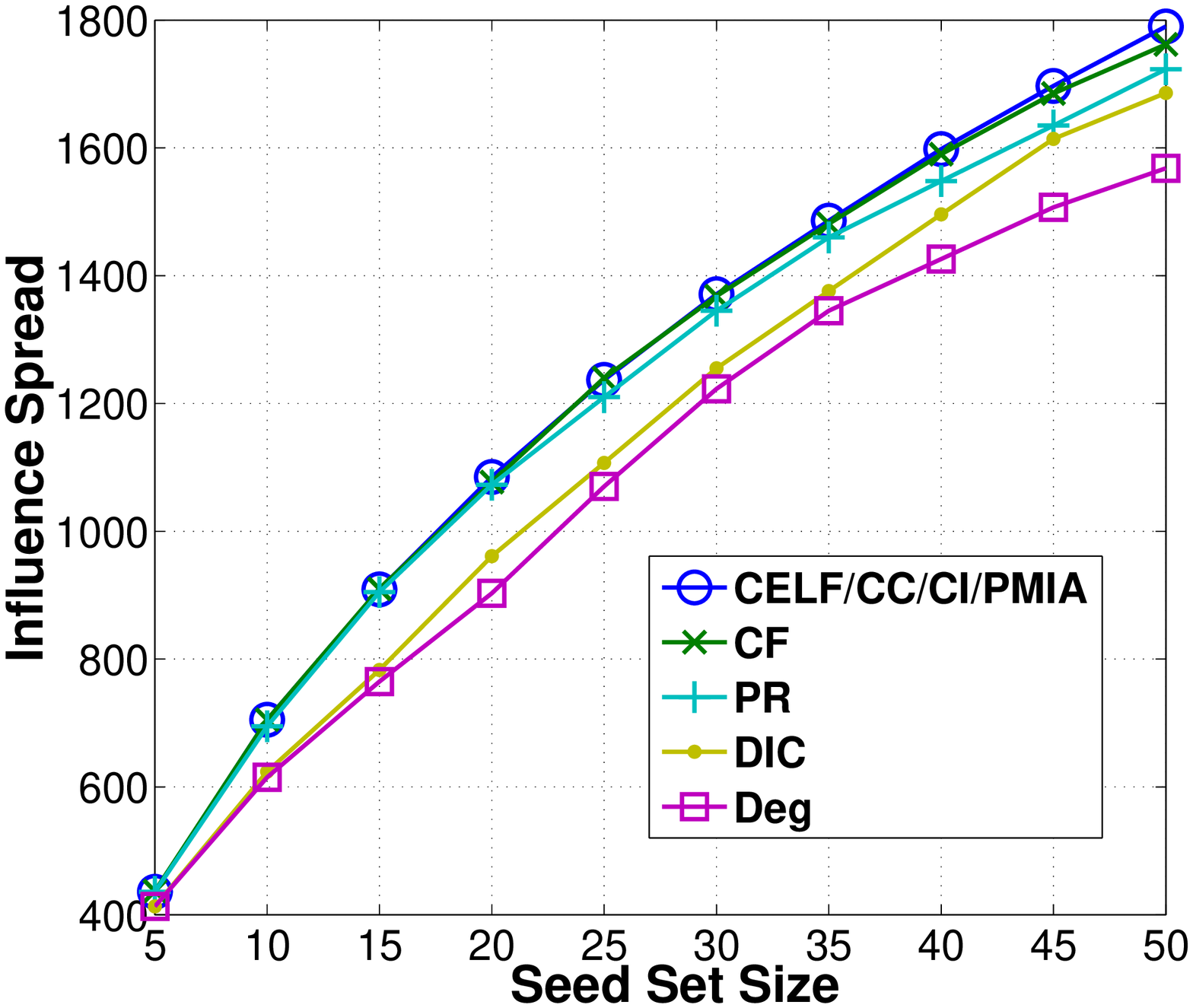}}   
    \subfigure[  ca-HepPh.
    \quad]{\label{fig:amazon}\includegraphics[scale=0.25]{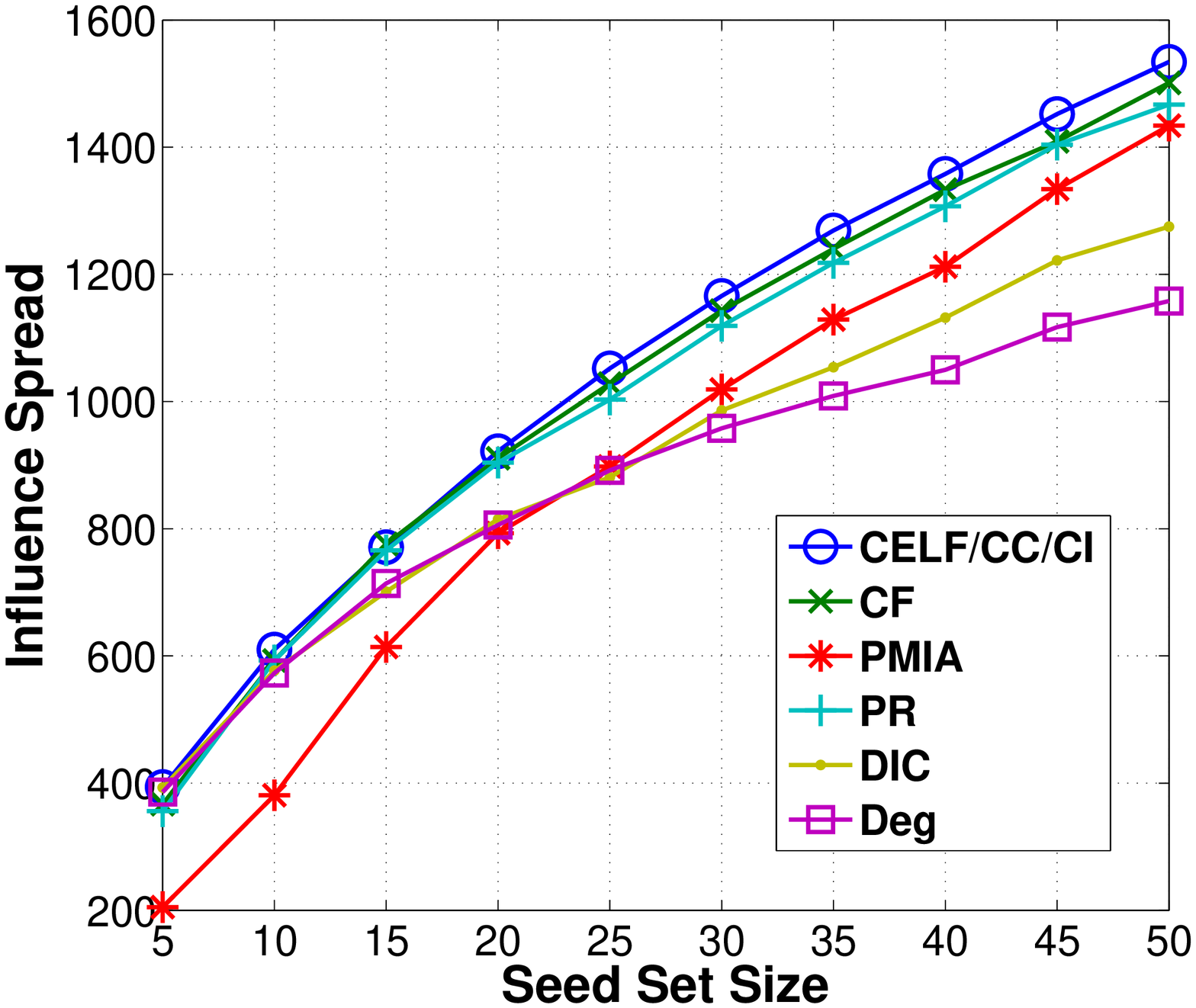}}  
    \subfigure[  web-NotreDame.
    \quad]{\label{fig:wiki}\includegraphics[scale=0.25]{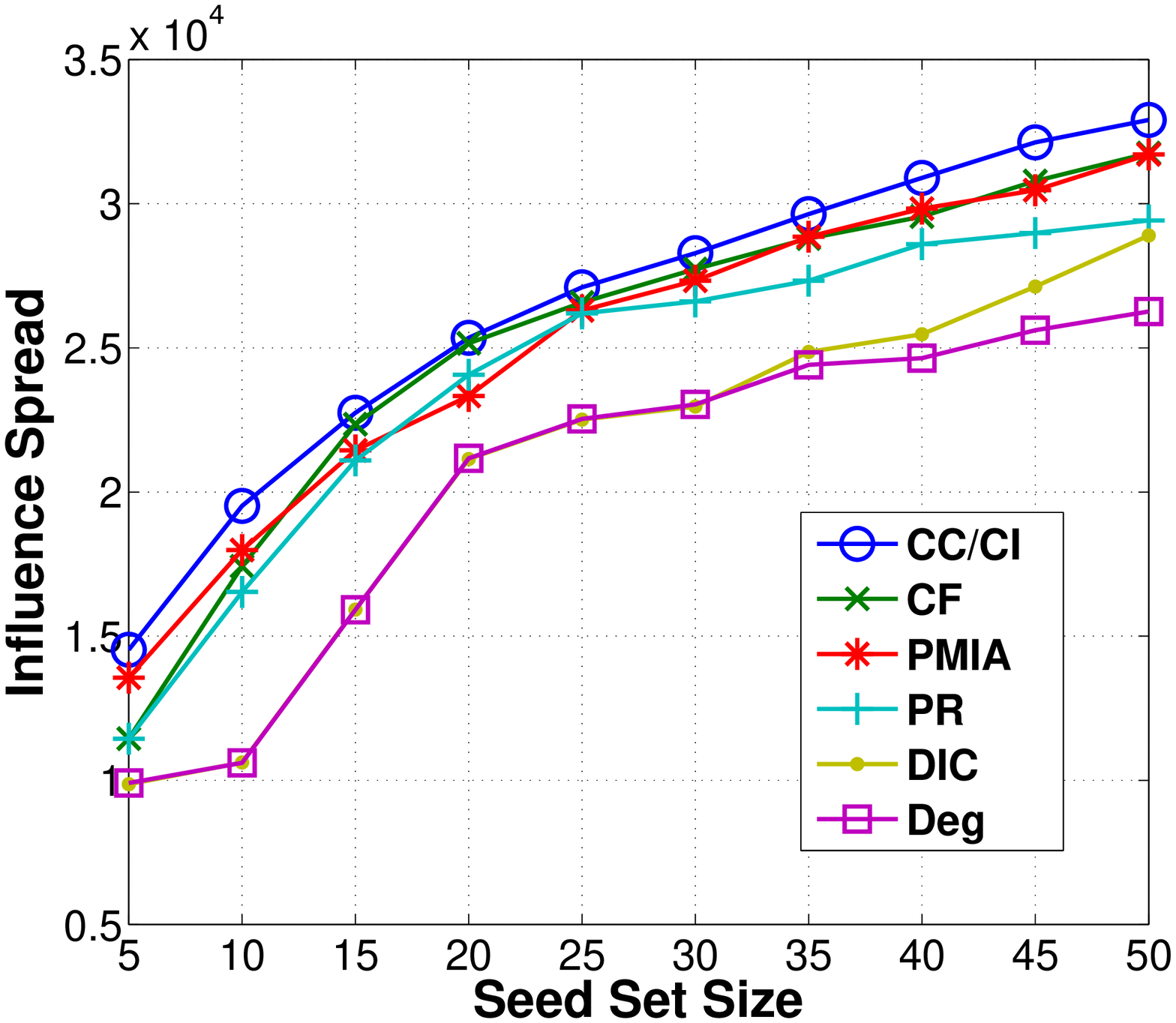}}  
    \subfigure[ DBLP.
    \quad]{\label{fig:phy}\includegraphics[scale=0.25]{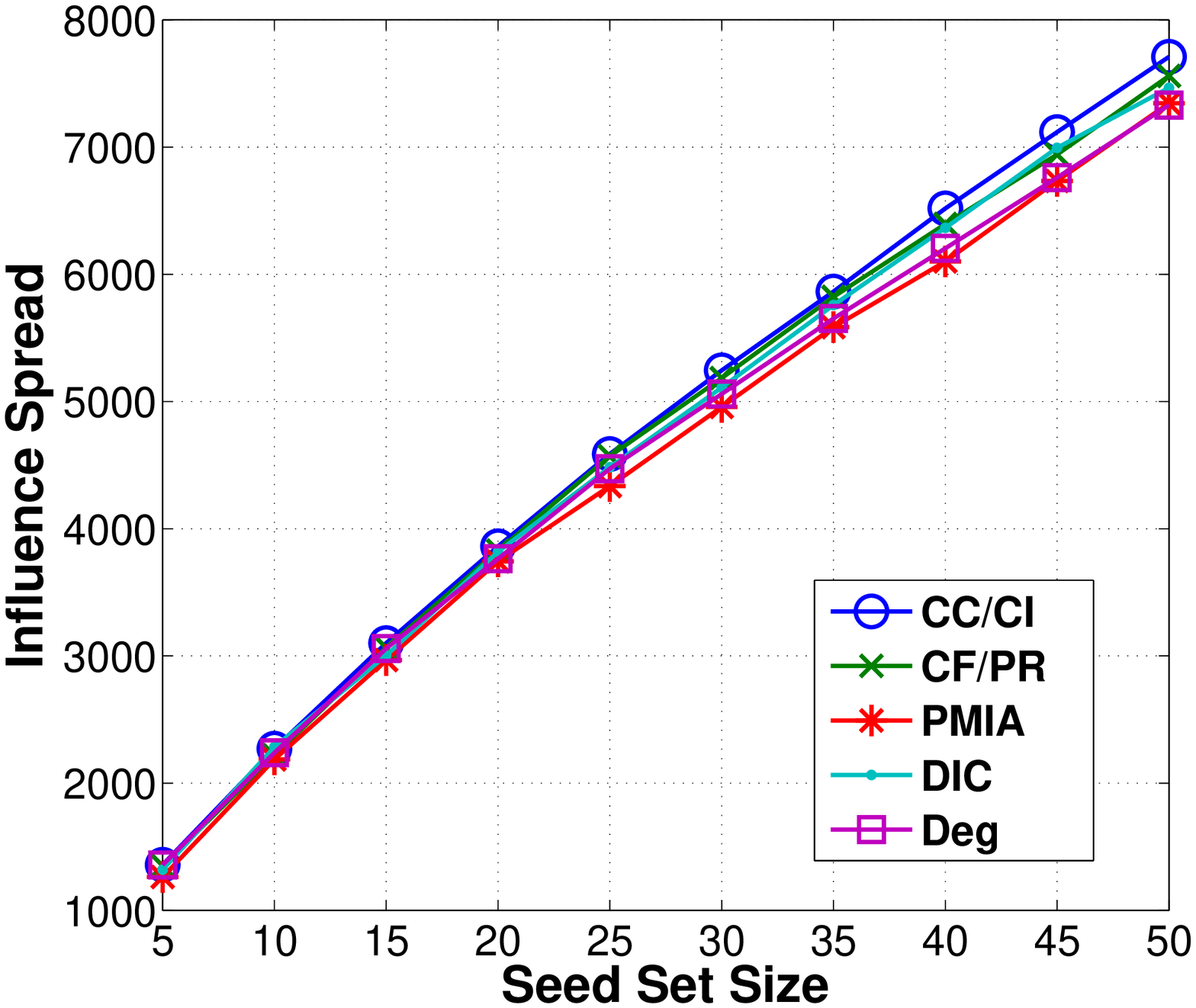}}   
    \subfigure[  LiveJournal.
    \quad]{\label{fig:amazon}\includegraphics[scale=0.25]{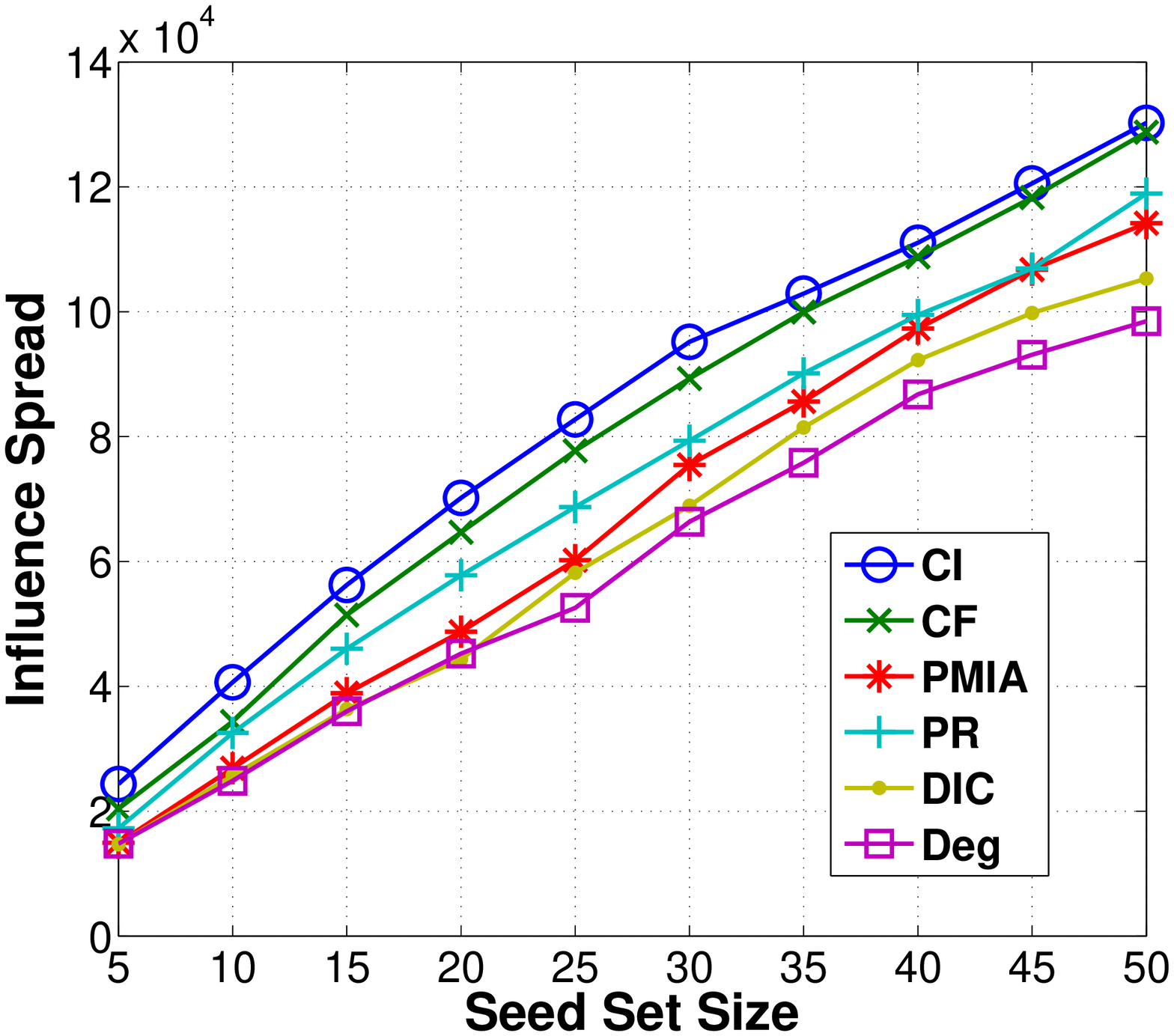}}  
  \end{center}
    \caption{The results of influence spread on six datasets. \label{fig:exp}}
\end{figure*}

\begin{figure*}
  \begin{center}\hspace{-1cm}
  \includegraphics[scale=0.85]{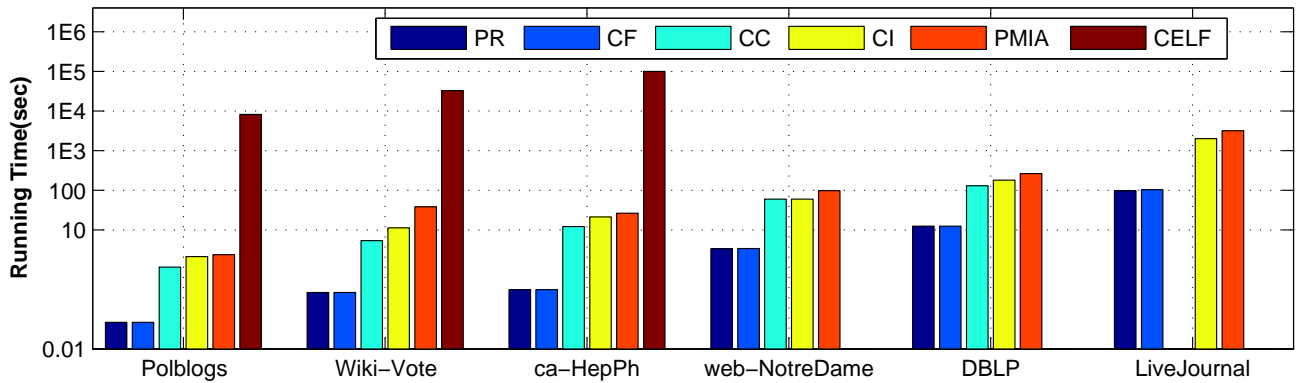} \caption{The computational performances. \label{fig:time}}
  \end{center}
\end{figure*}

In the following, we present a performance comparison of both
effectiveness and efficiency between our algorithms in this paper and the benchmarks. For
the purpose of comparison, we record the best performance of each
algorithm by tuning their parameters. We run tests on the six
networks under the WC model to obtain the results of influence spread. The seed
set size $K$ ranges from 1 to 50. Figure~\ref{fig:exp} shows the
final results of influence spread, where we paint tokens at
each 5 points. In this figure, if two curves are too close to each other, we group them
together and show properly in the legend. Figure~\ref{fig:time} shows the computational performance comparison for selecting 50 seeds on the best parameters. In this figure, for the running time of \textbf{DIC} and \textbf{Deg} is almost 0, we just remove their performances on this figure. Due to the running time overflow, \textbf{CELF} is failure in networks \textbf{web-NotreDame}, \textbf{DBLP}, \textbf{LiveJournal}. Due to the memory overflow, \textbf{CC} is failure in network \textbf{LiveJournal}.

From Figure~\ref{fig:exp}, we could get a beat record for each pair of algorithms. For example, if algorithm $A$ beat algorithm $B$ in $x$ datasets, then we could get a record $(A, B, x)$. Then we put these records into a table where the value in $(A,B)$-entry is $x$. Table~\ref{tab:beattable} is the beat table. Moreover, the last column of Table~\ref{tab:beattable} shows the total number of beat times, and the last row shows the total number of defeated times. For an algorithm, if we use the difference between its total number of beat times and defeated times as its strength, we could get its position in all of algorithms. The differences of the seven algorithms are 25, 29, 10, -5, -5, -20, -34 respectively. Based on this number, we could get the order of these algorithms, i.e., $\textbf{CI}>\textbf{CC}>\textbf{CF}>\textbf{PMIA}=\textbf{PR}>\textbf{DIC}>\textbf{Deg}$ where ``$>$'' means ``is better than''. However, actually, the performance of \textbf{CC} is even a little better than \textbf{CI} in five networks excluding LiveJournal. Because of its failure in \textbf{LiveJournal}, its overall performance is worse than \textbf{CI}. Besides, we didn't listed \textbf{CELF} in the beat table for it only succeed in three networks. But its performance is a little better than \textbf{CC}. 

In aspect of running time, we illustrated the computational costs of different algorithms on different datasets in Figure~\ref{fig:time}. For the running time of \textbf{DIC} and \textbf{Deg} are almost equal to 0, we just removed them from the figure. We could see that, in this aspect, the order of algorithm is $\textbf{PR}>\textbf{CF}>\textbf{CC}>\textbf{CI}>\textbf{PMIA}>\textbf{CELF}$ where ``$>$'' means ``is faster than''. Notably, the running time of \textbf{CF} algorithm is almost equal to $\textbf{PR}$ which means that \textbf{CF} is a linear time algorithm for viral marketing. Based on the discussion in Section~\ref{subsec:authority}, we know that the authority of individual is essentially her total influence in the network. Thus, \textbf{PR} could find the top-K most influential individuals. However these individuals may overlap their influence field for there are no mechanism to guarantee that they all have exclusive territories. While \textbf{CF} could guarantee it in some way and then it could always beat \textbf{PR}.

\begin{table}\label{tab:beattable}
\caption{The Beat Table.}
\begin{center}
 \begin{tabular}{|l|c|c|c|c|c|c|c|c|}
  \hline
        & CC & CI & CF & PMIA & PR & DIC & Deg & Win \\ \hline\hline
  CC    & -  & 0  & 5  & 4    & 5  & 5   & 6   & 25    \\ \hline
  CI    & 0  & -  & 6  & 5    & 6  & 6   & 6   & 29    \\ \hline
  CF    & 0  & 0  & -  & 4    & 6  & 6   & 6   & 22    \\ \hline
  PMIA  & 0  & 0  & 1  & -    & 3  & 4   & 5   & 13    \\ \hline
  PR    & 0  & 0  & 0  & 3    & -  & 6   & 6   & 15    \\ \hline
  DIC   & 0  & 0  & 0  & 1    & 0  & -   & 6   & 7    \\ \hline
  Deg   & 0  & 0  & 0  & 1    & 0  & 0   & -   & 1    \\ \hline
  Loss  & 0  & 0  & 12 & 18   & 20 & 27  & 35  &       \\ \hline
  \hline
\end{tabular}   
\end{center} 
\end{table}

\begin{figure*}
  \begin{center}
    \subfigure[  Polblogs.
    \quad]{\label{fig:wiki}\includegraphics[scale=0.25]{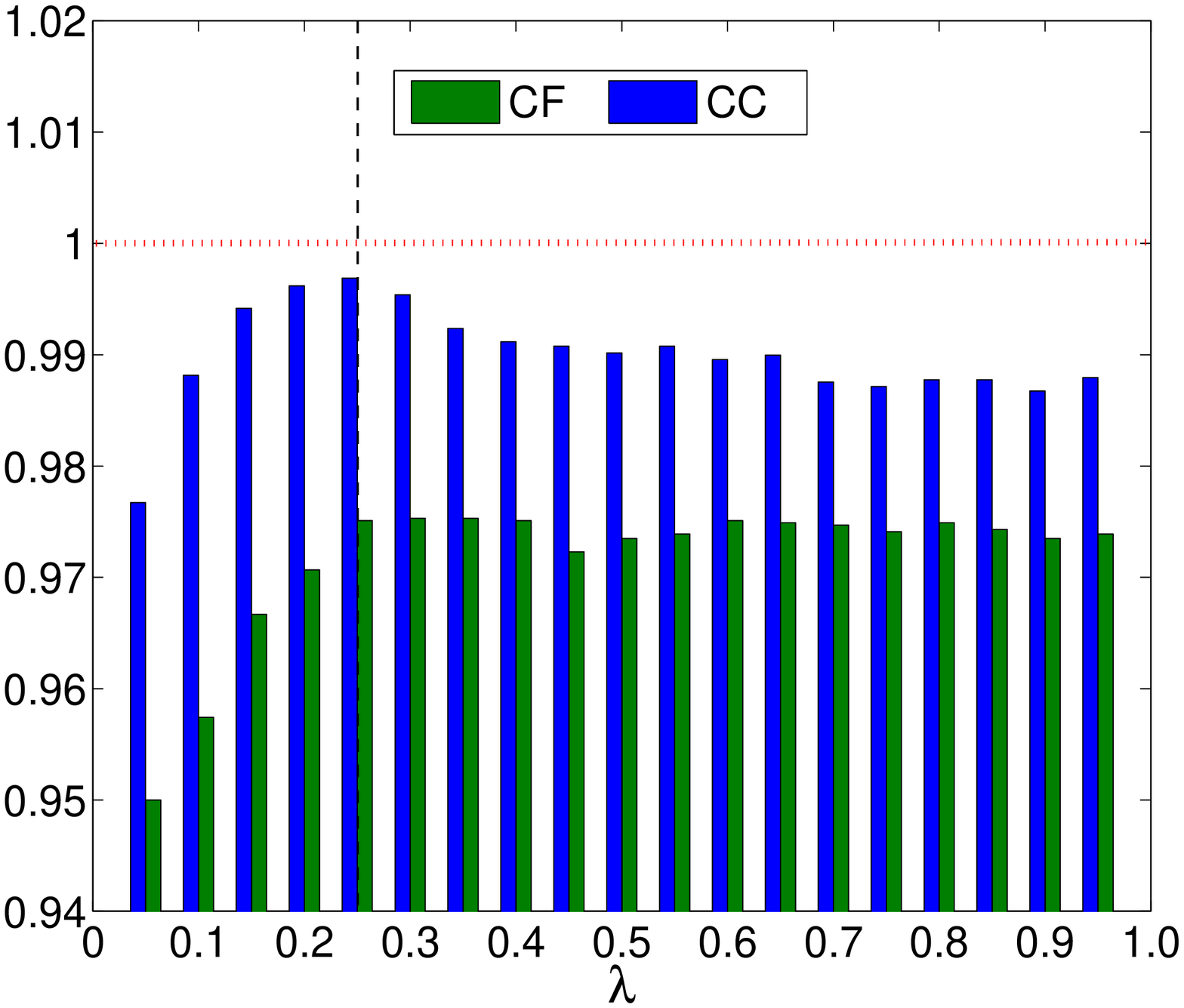}}  
    \subfigure[  Wiki-Vote.
    \quad]{\label{fig:amazon}\includegraphics[scale=0.25]{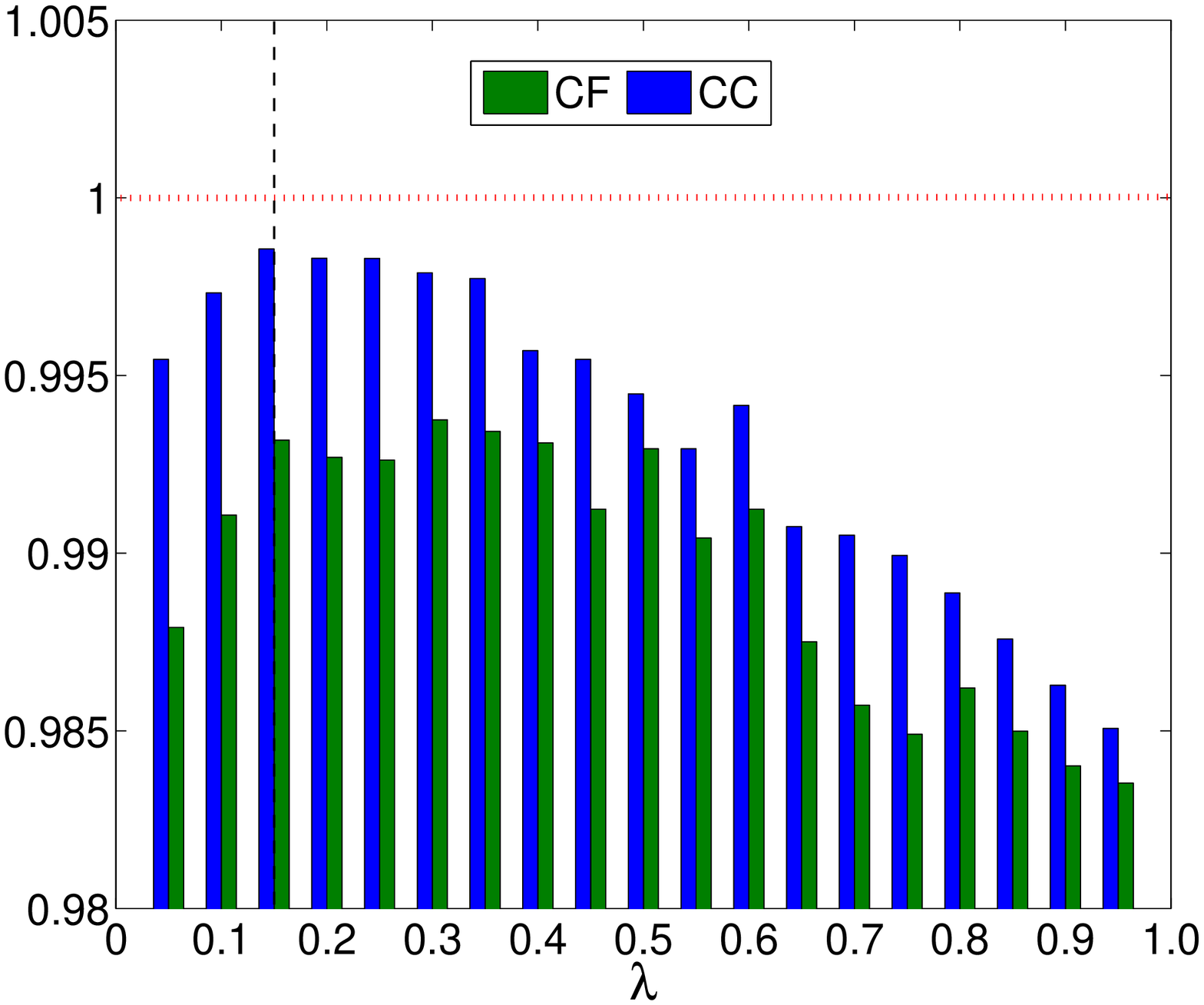}}  
    \subfigure[ ca-HepPh.
    \quad]{\label{fig:phy}\includegraphics[scale=0.25]{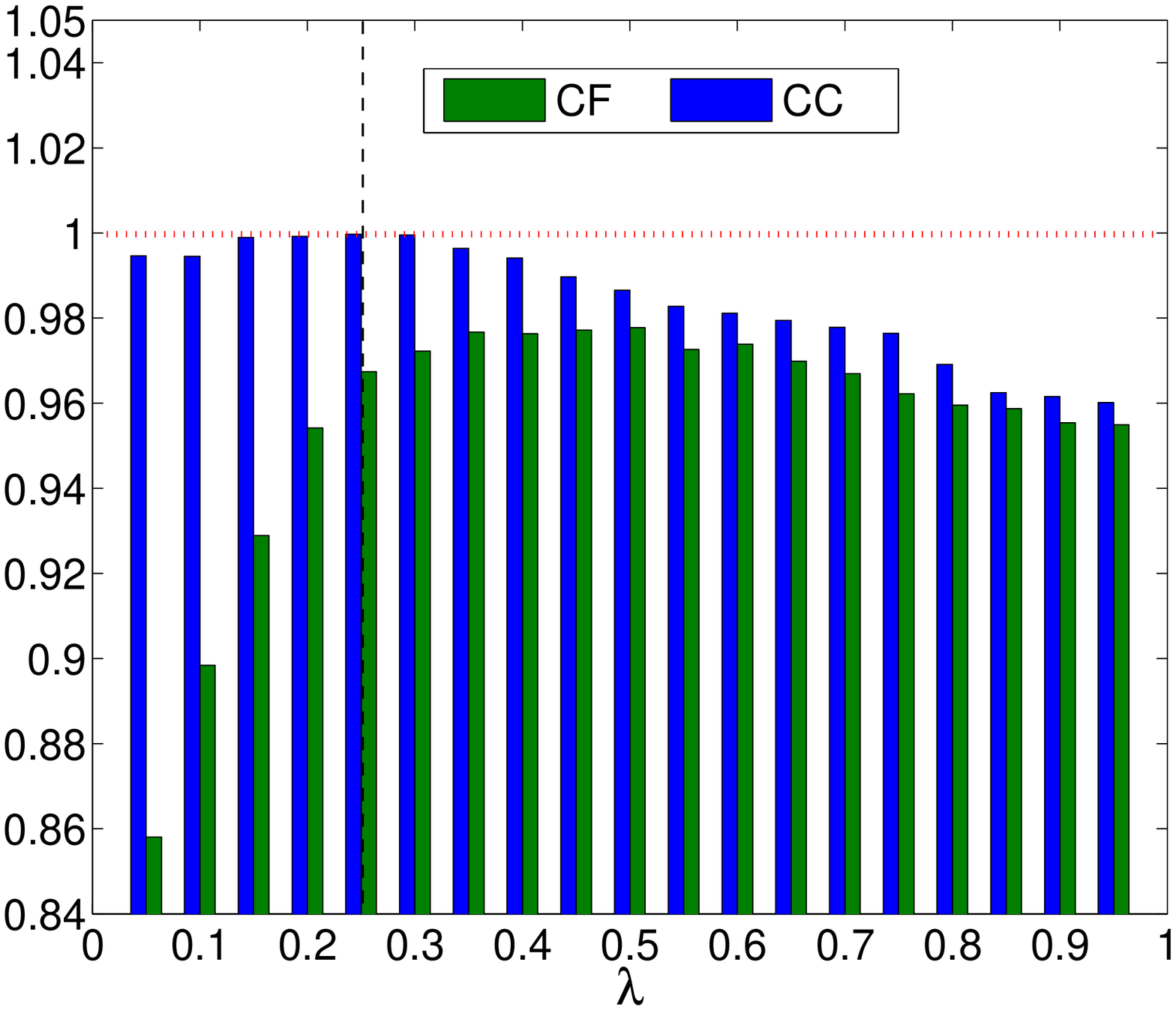}}   
    \subfigure[  Polblogs.
    \quad]{\label{fig:phy}\includegraphics[scale=0.25]{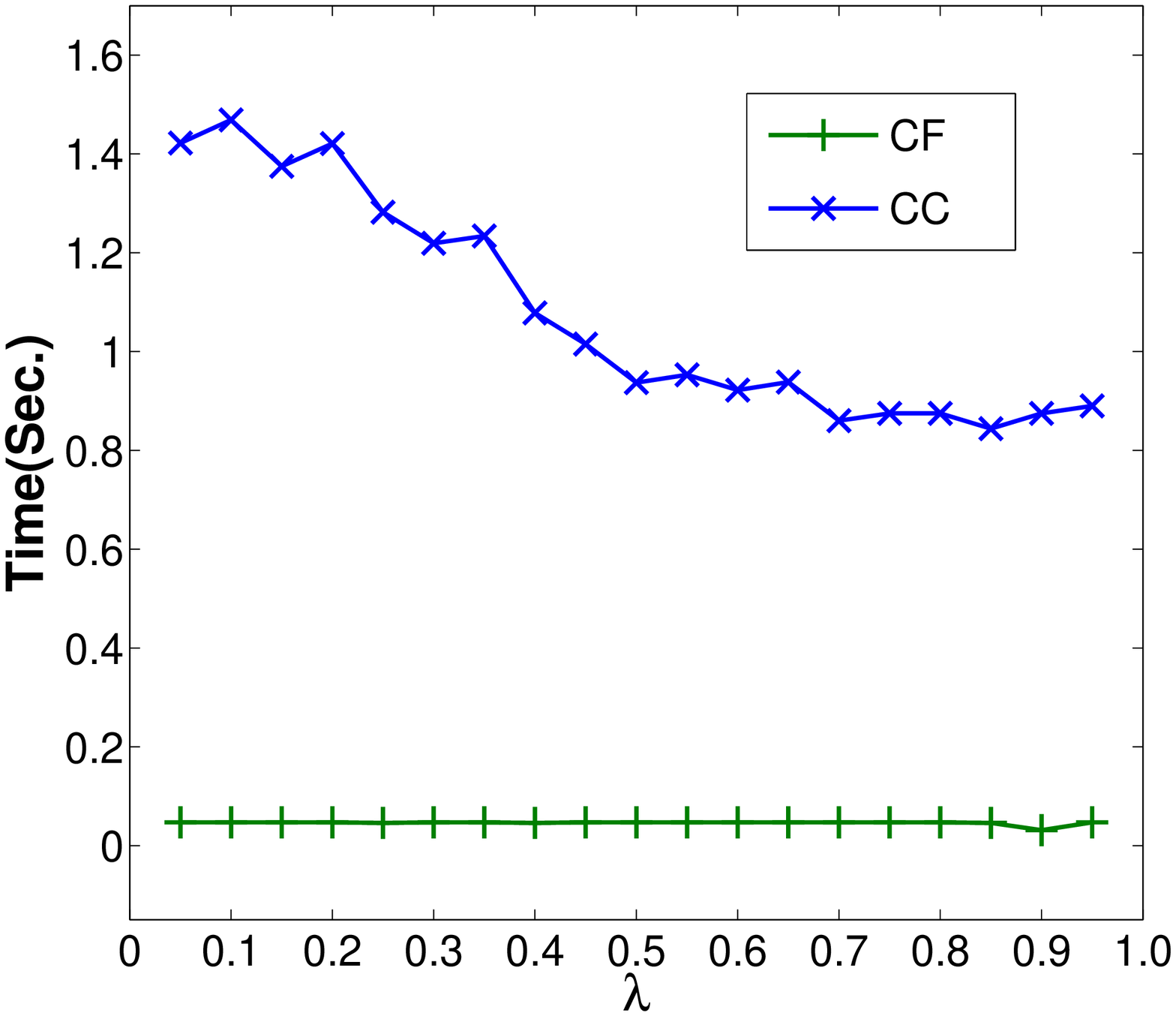}}   
    \subfigure[  Wiki-Vote.
    \quad]{\label{fig:wiki}\includegraphics[scale=0.25]{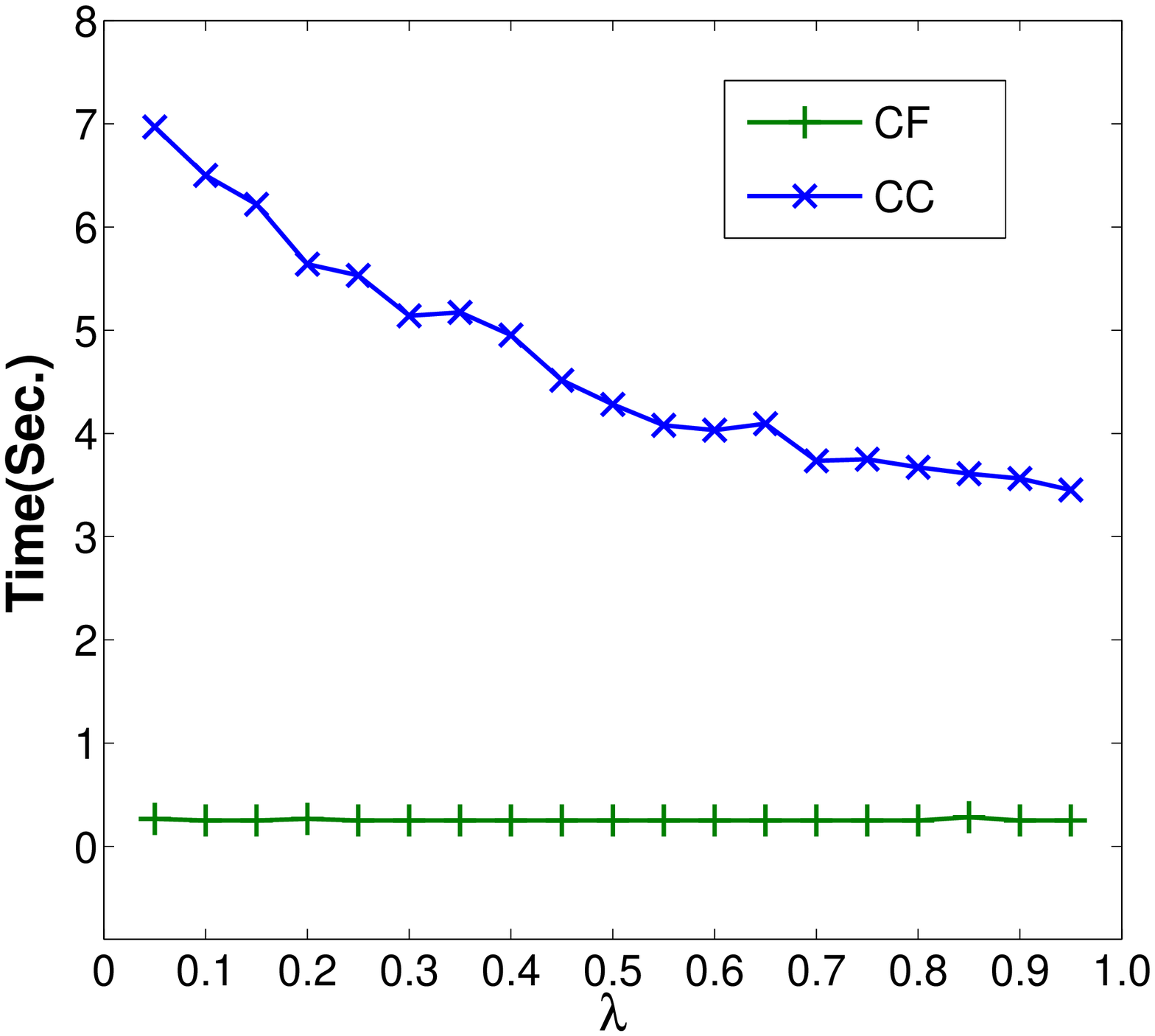}}  
    \subfigure[  ca-HepPh.
    \quad]{\label{fig:amazon}\includegraphics[scale=0.25]{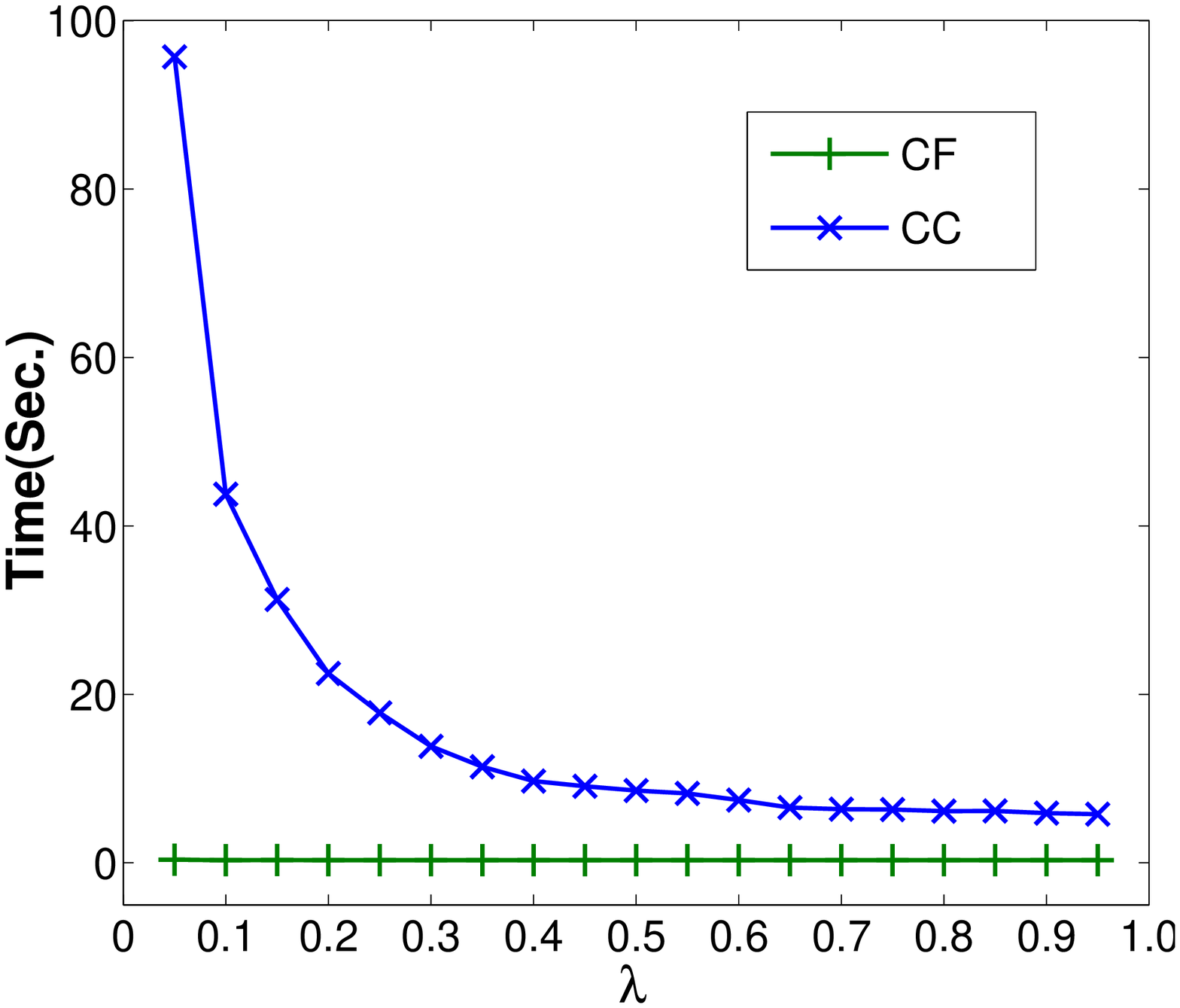}}  
  \end{center}
    \begin{center}
   \caption{The variation of effectiveness and the running time of \textbf{CF} and \textbf{CC} as the change of $\lambda$ on three network datasets: Polblogs, Wiki-Vote, and ca-HepPh. The top row of figures show the variation of effectiveness and the bottom row of figures show the variation of the  running time. \label{fig:lambda}}
  \end{center}
\end{figure*}

\textbf{Summary.} Generally, for solving the viral marketing problem, \textbf{CC} and \textbf{CI} perform consistently well on
each network, when the network size is large-scale, \textbf{CI} is a more proper choice. If we want to adopt a more faster algorithm, \textbf{CF} is the best. 

\subsubsection{The Impact of $\lambda$}

We investigate the effect of tuning parameter $\lambda$ on the
running time of \textbf{CC} and \textbf{CF} and the results of its influence spread.
Specifically, we set $\lambda$ ranges from 0.05 to 1, step by 0.05,
and then get the corresponding influence spread and running time.
And, for a clear view of the influence spread results, we use the
ratio of their influence spread result relative to CELF's to
indicate their effectiveness.

The up row of Figure~\ref{fig:lambda}
show the effectiveness of \textbf{CC} and \textbf{CF} with different $\lambda$ on Polblogs, Wiki-Vote, ca-HepPh network respectively. In these figures, the x axis
is the $\lambda$ value; the red dash line is $y=1$ which
indicates the results of CELF; the black dash line indicates the optimal value. From these figures, we can obtain the following observations:
\begin{itemize}
  \item The performances of  \textbf{CC} and \textbf{CF} all increased firstly and then decreased which follows the same trend appeared in Figure~\ref{fig:modelcomp}, but the optimal value is reached at $\lambda=0.25,0.15,0.25$ respectively; they are all reached a little later than the peak values in Figure~\ref{fig:modelcomp}; 
  \item The performance of of  \textbf{CC} is very stable. No matter what value $\lambda$ is, the difference of effectiveness is less than 0.04, and for most of $\lambda$ values, the effectiveness of \textbf{CC} is larger than 0.98.
  \item The best $\lambda$ located in the range $[0.1,0.4]$.
\end{itemize}

The bottom row of Figure~\ref{fig:lambda}
show the running time of  \textbf{CC} and \textbf{CF} with different $\lambda$ on Polblogs,  Wiki-Vote, ca-HepPh
respectively.  On these figures, we can observe that the running
time of \textbf{CC} is descending with the ascending of $\lambda$ while the running time of \textbf{CF} always stays at a constant value. From the above
observations, we can know that,  for \textbf{CC}, if we want to get a better
effectiveness we should set $\lambda$ to be a number in
$[0.1,0.4]$ and if we want to get the result efficiently, we should set
$\lambda$ to be a comparable large value; while for \textbf{CF}, we just directly set $\lambda$ to be a number in $[0.1,0.4]$.

\section{Conclusion}\label{sec:conclusion}
In this paper, we developed a social influence model based on circuit theory for describing the information propagation in social networks. This model is tractable and flexible for understanding patterns of information propagation. Under this model, several upper bound properties were identified. These properties can help us to quickly locate the nodes to be considered during the information propagation process. This can drastically reduce the search space, and thus vastly improve the efficiency of measuring the influence strength between any pair of nodes. In addition, the circuit theory based model provides a new way to compute the independent influence of nodes and leads to a natural solution to the social influence maximization problem. Finally, experimental results showed the advantages of the circuit theory based model over the existing models in terms of efficiency as well as the effectiveness for measuring the information propagation in social networks.

\bibliographystyle{abbrv}

\begin{thebibliography}{10}

\bibitem{adamic2005political}
L.~Adamic and N.~Glance.
\newblock The political blogosphere and the 2004 us election: divided they
  blog.
\newblock In {\em Proceedings of the 3rd international workshop on Link
  discovery}, pages 36--43. ACM, 2005.

\bibitem{aggarwal2011social}
C.~Aggarwal.
\newblock {\em Social network data analytics}.
\newblock Springer-Verlag New York Inc, 2011.

\bibitem{aggarwalflow}
C.~Aggarwal, A.~Khan, and X.~Yan.
\newblock On flow authority discovery in social networks.
\newblock In {\em Proceedings of the SDM Conference}, 2011.

\bibitem{anagnostopoulos2008influence}
A.~Anagnostopoulos, R.~Kumar, and M.~Mahdian.
\newblock Influence and correlation in social networks.
\newblock In {\em Proceeding of the 14th ACM SIGKDD international conference on
  Knowledge discovery and data mining}, pages 7--15. ACM, 2008.

\bibitem{bass2004new}
F.~Bass.
\newblock A new product growth for model consumer durables.
\newblock {\em MANAGEMENT SCIENCE}, 50(12 Supplement):1825--1832, 2004.

\bibitem{bianchini2005inside}
M.~BIANCHINI, M.~GORI, and F.~SCARSELLI.
\newblock Inside pagerank.
\newblock {\em ACM Transactions on Internet Technology}, 5(1):92--128, 2005.

\bibitem{brown1987social}
J.~Brown and P.~Reingen.
\newblock Social ties and word-of-mouth referral behavior.
\newblock {\em Journal of Consumer Research}, pages 350--362, 1987.

\bibitem{chen2010scalable}
W.~Chen, C.~Wang, and Y.~Wang.
\newblock Scalable influence maximization for prevalent viral marketing in
  large-scale social networks.
\newblock In {\em Proceedings of the 16th ACM SIGKDD international conference
  on Knowledge discovery and data mining}, pages 1029--1038. ACM, 2010.

\bibitem{chen2009efficient}
W.~Chen, Y.~Wang, and S.~Yang.
\newblock Efficient influence maximization in social networks.
\newblock In {\em Proceedings of the 15th ACM SIGKDD international conference
  on Knowledge discovery and data mining}, pages 199--208. ACM, 2009.

\bibitem{domingos2001mining}
P.~Domingos and M.~Richardson.
\newblock Mining the network value of customers.
\newblock In {\em Proceedings of the seventh ACM SIGKDD international
  conference on Knowledge discovery and data mining}, pages 57--66. ACM, 2001.

\bibitem{easley2010networks}
D.~Easley and J.~Kleinberg.
\newblock {\em Networks, crowds, and markets: Reasoning about a highly
  connected world}.
\newblock Cambridge Univ Pr, 2010.

\bibitem{goldenberg2001talk}
J.~Goldenberg, B.~Libai, and E.~Muller.
\newblock Talk of the network: A complex systems look at the underlying process
  of word-of-mouth.
\newblock {\em Marketing letters}, 12(3):211--223, 2001.

\bibitem{goldenberg2001using}
J.~Goldenberg, B.~Libai, and E.~Muller.
\newblock Using complex systems analysis to advance marketing theory
  development: Modeling heterogeneity effects on new product growth through
  stochastic cellular automata.
\newblock {\em Academy of Marketing Science Review}, 9(3):1--18, 2001.

\bibitem{goyal2010learning}
A.~Goyal, F.~Bonchi, and L.~Lakshmanan.
\newblock Learning influence probabilities in social networks.
\newblock In {\em Proceedings of the third ACM international conference on Web
  search and data mining}, pages 241--250. ACM, 2010.

\bibitem{granovetter1978threshold}
M.~Granovetter.
\newblock Threshold models of collective behavior.
\newblock {\em American journal of sociology}, pages 1420--1443, 1978.

\bibitem{kempe2003maximizing}
D.~Kempe, J.~Kleinberg, and {\'E}.~Tardos.
\newblock Maximizing the spread of influence through a social network.
\newblock In {\em Proceedings of the ninth ACM SIGKDD international conference
  on Knowledge discovery and data mining}, pages 137--146. ACM, 2003.

\bibitem{kimura2006tractable}
M.~Kimura and K.~Saito.
\newblock Tractable models for information diffusion in social networks.
\newblock {\em Knowledge Discovery in Databases: PKDD 2006}, pages 259--271,
  2006.

\bibitem{kirchhoff}
G.~Kirchhoff.
\newblock Vorlesungen ueber mathematische physik, mechanik.
\newblock {\em Leipzig: Teubner}, 1877.

\bibitem{leskovec2007cost}
J.~Leskovec, A.~Krause, C.~Guestrin, C.~Faloutsos, J.~VanBriesen, and
  N.~Glance.
\newblock Cost-effective outbreak detection in networks.
\newblock In {\em Proceedings of the 13th ACM SIGKDD international conference
  on Knowledge discovery and data mining}, pages 420--429. ACM, 2007.

\bibitem{mahajan1990new}
V.~Mahajan, E.~Muller, and F.~Bass.
\newblock New product diffusion models in marketing: A review and directions
  for research.
\newblock {\em The Journal of Marketing}, pages 1--26, 1990.

\bibitem{page1999pagerank}
L.~Page, S.~Brin, R.~Motwani, and T.~Winograd.
\newblock The pagerank citation ranking: Bringing order to the web.
\newblock In {\em Proceedings of WWW Conference}. Stanford InfoLab, 1999.

\bibitem{richardson2002mining}
M.~Richardson and P.~Domingos.
\newblock Mining knowledge-sharing sites for viral marketing.
\newblock In {\em Proceedings of the eighth ACM SIGKDD international conference
  on Knowledge discovery and data mining}, pages 61--70. ACM, 2002.

\bibitem{wang2010community}
Y.~Wang, G.~Cong, G.~Song, and K.~Xie.
\newblock Community-based greedy algorithm for mining top-k influential nodes
  in mobile social networks.
\newblock In {\em Proceedings of the 16th ACM SIGKDD international conference
  on Knowledge discovery and data mining}, pages 1039--1048. ACM, 2010.

\bibitem{biao2012linear}
B.~Xiang, E.~Chen, Q.~Liu, H.~Xiong, Y.~Yang, and J.~Xie.
\newblock A circuit inspired linear model for social influence analysis.
\newblock In {\em Proceedings of the IEEE 11th International Conference on Data
  Mining (ICDM)}, pages xxx--xxx. IEEE, 2012.

\bibitem{yang2012approximation}
Y.~Yang, E.~Chen, Q.~Liu, B.~Xiang, T.~Xu, and S.~Shad.
\newblock On approximation of real-world inuencespread.
\newblock In {\em The European Conference on Machine Learning and Principles
  and Practice of Knowledge Discovery in Databases (PKDD)}, pages xxx--xxx,
  2012.

\bibitem{Zafarani+Liu:2009}
R.~Zafarani and H.~Liu.
\newblock Social computing data repository at {ASU}, 2009.

\end{thebibliography}

\end{document}